\documentclass{aa}
\usepackage{mathabx}
\usepackage[title]{appendix}
\usepackage{natbib}
\usepackage{color}
\usepackage{hyperref}
\usepackage{academicons}
\usepackage{xcolor}

\usepackage{graphicx}
\usepackage{txfonts}
\begin{document}

   \title{\texttt{Wapiti}: a data-driven approach to correct for systematics in RV data}
   \subtitle{Application to SPIRou data of the planet-hosting M dwarf GJ\,251}
   \titlerunning{The \texttt{Wapiti} method - Application to SPIRou data of the planet-hosting M dwarf GJ\,251}

   \author{M. Ould-Elhkim \inst{1}         
              \fnmsep\thanks{Based on observations obtained at the Canada-France-Hawaii Telescope (CFHT) which is 
              operated from the summit of Maunakea by the National Research Council of Canada, the Institut National des Sciences de l'Univers of the Centre 
              National de la Recherche Scientifique of France, and the University of Hawaii. The observations at the Canada-France-Hawaii Telescope were 
              performed with care and respect from the summit of Maunakea which is a significant cultural and historic site.
              Based on observations obtained with SPIRou, an international project led by Institut de Recherche en Astrophysique et Plan\'etologie, Toulouse, France.
              SPIRou is an acronym for SPectropolarimetre InfraROUge (infrared spectropolarimeter).}
          \and
          C. Moutou\inst{1}
          \and
          J-F. Donati\inst{1}
          \and
          \'E. Artigau\inst{2,3}
          \and
          P. Fouqué\inst{1}
          \and
          N.J. Cook\inst{2}
          \and
          A. Carmona\inst{4}
          \and
          P. I. Cristofari\inst{1,5}
          \and
          E. Martioli\inst{6, 7}
          \and
          F. Debras\inst{1}
          \and
          X. Dumusque\inst{8}
          \and
          J. H. C. Martins\inst{10}
          \and
          G. Hébrard\inst{6}
          \and
          C. Cadieux \inst{2}
          \and
          X. Delfosse\inst{4}
          \and
          R. Doyon\inst{2}
          \and
          B. Klein\inst{9}
          \and
          J. Gomes da Silva \inst{10}
          \and
          T. Forveille\inst{4}
          \and
          T. Hood\inst{1} 
          \and
          P. Charpentier\inst{1}
          }

    \institute{Universit\'e de Toulouse, UPS-OMP, IRAP, 14 avenue E. Belin, Toulouse, F-31400, France, France\\
    \email{merwan.ould-elhkim@irap.omp.eu}     
    \and  
    Trottier Institute for Research on Exoplanets, Université de Montréal, Département de Physique, C.P. 6128 Succ. Centre-ville, Montréal,
    QC H3C 3J7, Canada
    \and 
    Observatoire du Mont-M\'egantic, Universit\'e de Montr\'eal, D\'epartement de Physique, C.P. 6128 Succ. Centre-ville, Montr\'eal, QC H3C 3J7, Canada
    \and 
      Univ. Grenoble Alpes, CNRS, IPAG, 38000 Grenoble, France
    \and 
   Center for Astrophysics | Harvard \& Smithsonian, 60 Garden Street, Cambridge, MA 02138, USA
    \and 
    Institut d'Astrophysique de Paris, CNRS, UMR 7095, Sorbonne Universit\'{e}, 98 bis bd Arago, 75014 Paris, France
     \and 
     Laborat\'{o}rio Nacional de Astrof\'{i}sica, Rua Estados Unidos 154, 37504-364, Itajub\'{a} - MG, Brazil 
    \and     
     D\'epartement d’astronomie, Universit\'e de Gen\`eve, Chemin des Maillettes 51, CH-1290 Versoix, Switzerland
    \and
    Department of Physics, University of Oxford, Oxford OX1 3RH, UK
    \and
    Instituto de Astrofísica e Ciências do Espaço, Universidade do Porto, CAUP, Rua das Estrelas, 4150-762 Porto, Portugal\\
     }

 
  \abstract
   {Recent advances in the development of precise radial velocity (RV) instruments in the near-infrared (nIR) domain, such as SPIRou, have facilitated the study of M-type stars to more effectively characterize planetary systems. However, the nIR presents unique challenges in exoplanet detection due to various sources of planet-independent signals which can result in systematic errors in the RV data.}
   {In order to address the  challenges posed by the detection of exoplanetary systems around M-type stars using nIR observations, we introduce a new data-driven approach for correcting systematic errors in RV data. The effectiveness of this method is demonstrated through its application to the star GJ\,251.}
   {Our proposed method, referred to as \texttt{Wapiti} (Weighted principAl comPonent analysIs reconsTructIon), uses a dataset of per-line RV time-series generated by the line-by-line (LBL) algorithm and employs a weighted principal component analysis (wPCA) to reconstruct the original RV time-series. A multi-step process is employed to determine the appropriate number of components, with the ultimate goal of subtracting the wPCA reconstruction of the per-line RV time-series from the original data in order to correct systematic errors.} 
   {The application of \texttt{Wapiti} to GJ\,251 successfully eliminates spurious signals from the RV time-series and enables the first detection in the nIR of GJ\,251b, a known temperate super-Earth with an orbital period of 14.2 days. This demonstrates that, even when systematics in SPIRou data are unidentified, it is still possible to effectively address them and fully realize the instrument's capability for exoplanet detection. Additionally, in contrast to the use of optical RVs, this detection did not require to filter out stellar activity, highlighting a key advantage of nIR RV measurements.}
   {}

   \keywords{ Techniques: radial velocities;  Techniques: spectroscopic;  Methods: data analysis; Stars: individual: GJ\,251; Stars: planetary systems; Stars: low-mass
               }

   \maketitle
%

\section{Introduction}

Over the past few years, the development of precise radial velocity (RV) instruments in the near-infrared (nIR) domain, such as HPF \citep{2012SPIE.8446E..1SM}, IRD \citep{2014SPIE.9147E..14K}, CARMENES-NIR \cite{2017MS&E..278a2191B}, NIRPS \citep{2018EPSC...12.1147B} and SPIRou \citep{2020MNRAS.498.5684D}, has facilitated the detection and characterization of new exoplanetary systems \citep{2022AJ....164...96C, 2022A&A...660A..86M}. These instruments have enabled the expansion of research on M-type stars to the $YJHK$ spectral bandpasses \citep{quirrenbach_carmenes_2018, 2020ApJS..247...11R}. Those stars are of particular interest because they constitute the majority of main sequence stars in the solar neighborhood \citep{2006AJ....132.2360H}. Moreover, thanks to their low masses, M dwarfs are particularly well suited for the detection of planet RV signatures of a few m\,s$^{-1}$ \citep{2016Natur.536..437A, 2013A&A...549A.109B, Faria_2022, Su_rez_Mascare_o_2023} and they are known to harbor more rocky planets than stars of other types \citep{Bonfils_2013, DressingCharbonneau2015, 2016MNRAS.457.2877G, 2020MNRAS.498.2249H, 2021A&A...653A.114S}. 
\par 
However, M dwarfs pose challenges for exoplanet detection due to their high levels of magnetic activity, which produce spurious RV signals \citep{2007A&A...474..293B, Gomes_da_Silva_2012, 2016ApJ...821L..19N, 2016MNRAS.461.1465H}. Additonally, similar to other instruments operating in the optical range, some systematics may originate from the instrument itself, such as an average intra-night drift and a nightly zero-point variation that can be corrected \citep{2020A&A...636A..74T, 2019MNRAS.484L...8T}, or instrumental defects \citep{2015ApJ...808..171D}. Furthermore, the nIR range required to analyze the red spectral energy distribution of M dwarfs is prone to strong telluric absorption lines which can create an additional source of uncertainty in the measurement of radial velocity \citep{2014SPIE.9149E..05A}. The nIR poses other challenges, including persistence in infrared arrays. Most notably, this phenomenon is characterized by the fact that not only the preceding frame, but the entire illumination history spanning the last few hours, is of significance. Specifically, bright targets observed at the outset of a night can impact the observation of fainter targets later on in the same night \citep{2018SPIE10709E..1PA}.

\par 

Despite efforts to correct these systematics, it is possible that spurious signals may still remain after applying existing techniques, potentially due to incorrect assumptions, technical limitations, or unidentified sources. To address this issue, we present a new data-driven method for correcting systematics in RV data. We focus our work on data collected with the SPIRou instrument, a high-precision RV spectrograph in the nIR domain whose two forefront science topics: the quest for Earth-like planets in the habitable zones of very-low-mass stars, and the study of low-mass stars and planet formation in the presence of magnetic fields \citep{2020MNRAS.498.5684D}.

\par 

To achieve these science goals, SPIRou has a large program (LP) called the SPIRou Legacy Survey (SLS)\footnote{https://spirou-legacy.irap.omp.eu/doku.php} which was allocated 310 nights over 7 semesters (2019a to 2022a) on the Canada-France-Hawaii Telescope. The SLS-Planet Search program, which is a part of this LP, aims to detect and characterize planetary systems around $\sim$50 nearby M dwarfs (Moutou et al. in prep).

\par

The SPIRou data are prone to the previously mentioned systematics and it is necessary to develop methods to correct them. Our method, referred to as \texttt{Wapiti}\footnote{https://github.com/HkmMerwan/wapiti} (Weighted-principAl comPonent analysIs reconsTructIon), is agnostic to the origin of the systematics and reconstructs the per-line RV time-series. This method takes advantage of the new line-by-line (LBL) algorithm \citep{2022AJ....164...84A} used to compute RV data with SPIRou, which provides a RV time-series for each line of the spectrum. The application of PCA to mitigate systematics in RV data is a well-established technique in the exoplanet community. However, existing methods are primarily aimed at specifically reducing the impact of stellar activity, such as the methods presented in \citet{2021MNRAS.505.1699C} and \citet{2022A&A...659A..68C}. PCA based methods were also explored to correct for telluric or instrumental effects in prior studies \citep{2014SPIE.9149E..05A, 2021A&A...653A..43C}. Our method, on the other hand, is unique in that it does not focus solely on reducing the impact of stellar activity or telluric effects but instead is a versatile and flexible approach that can be applied to any target, as long as LBL data is available. We demonstrate the effectiveness of \texttt{Wapiti} by applying it to the star GJ\,251, a target with a known planet in the SLS-Planet Search program. 

\par 

This paper is structured as follows. Section \ref{sec:observations} provides an overview of the observations, data reduction, and LBL algorithm used for the study. Section \ref{sec:method} presents the \texttt{Wapiti} method in detail, including how it determines the optimal number of principal vectors. Section \ref{sec:results} demonstrates the efficiency of the \texttt{Wapiti} method by eliminating systematic peaks and confirming the detection of a planet with a 14.2-day period. Finally, section \ref{sec:ccl} summarizes the main results and discusses potential future work.


\section{Observations}\label{sec:observations}

\subsection{GJ\,251}

The star GJ\,251 (HD 265866) is a red dwarf with a spectral type of M3V \citep{1991ApJS...77..417K} located $5.5846 \pm 0.0009$ pc away \citep{2020yCat.1350....0G} from Earth. From the analysis of SPIRou spectra, \cite{2022MNRAS.516.3802C} derived the stellar parameters listed in Table \ref{gl251_stellar_properties} that we used in our study.

\begin{table}[h!]
\centering
\begin{tabular}{c  c}
\hline
\noalign{\smallskip}
Parameters & Value \\ [0.5ex]
\hline
\noalign{\smallskip}
\noalign{\smallskip}
$T_{\texttt{eff}}$ (K) & $3420 \pm 31$ \\     
\noalign{\smallskip}
$\left[M/H\right]$ & $-0.01 \pm 0.10$ \\
\noalign{\smallskip}
$\left[\alpha/\texttt{Fe}\right]$ & $-0.01 \pm 0.04$ \\
\noalign{\smallskip}
$\log\ g$ & $4.71 \pm 0.06$ \\
\noalign{\smallskip}
Radius $(R_\odot)$ & $0.365 \pm 0.007$ \\
\noalign{\smallskip}
Mass $(M_\odot)$ & $0.35 \pm 0.02$ \\
\noalign{\smallskip}
$\log \ (L/L_\odot)$ & $-1.786 \pm 0.003$ \\
\noalign{\smallskip}
\hline
\end{tabular}
\caption{Retrieved stellar parameters of GJ\,251 \citep{2022MNRAS.516.3802C}}
\label{gl251_stellar_properties}
\end{table}

In \citet{2017AJ....153..208B}, the detection of two planets with orbital periods of 1.74 and 601.32 days was claimed to be made using the HIRES spectrograph at the Keck observatory. However, in \citet{2020A&A...643A.112S}, no such evidence was found in the CARMENES visible data. Instead, a planet with an orbital period of $14.238 \pm 0.002$ days was identified in the data, with a semi-amplitude $K = 2.11 \pm 0.20 $ m\,s$^{-1}$, which falls within the category of temperate super-Earths, with a calculated mass of $M_p\sin i = 4.0 \pm 0.4\ M_\Earth $ and an equilibrium temperature of $T_{\texttt{eq}} = 351 \pm 1.4\ K$ \citep{2020A&A...643A.112S}. The small amplitude of this planet makes GJ\,251 an ideal candidate to demonstrate that SPIRou is also a suitable instrument for detecting Earth-like planets in the habitable zone of M dwarfs

\par

The observations were collected as series of circular polarization sequences of 4 individual sub exposures per night. The data collection for GJ,251 spans from December 12, 2018 to April 14, 2022, comprising a total of 769 sub-expositions, with each visit comprising four sub-exposures. Out of the available data, three visits are incomplete, leaving us with a total of 191 complete visits. The airmass during these observations varied from a minimum of 1.028 to a maximum of 2.003, with a median value of 1.083. The signal-to-noise ratio per 2.28-km\,s$^{-1}$ pixel bin in the middle of the $H$ band, as measured with 61-second exposures, has a median value of 137.0. 

\subsection{The SPIRou spectropolarimeter}

SPIRou is a high-resolution near-infrared spectropolarimeter installed in 2018 on the Canada-France-Hawaii Telescope (CFHT). It has a spectral range from 0.98 to 2.35 $\mu$m at a spectral resolving power of about 70\,000 \citep{2020MNRAS.498.5684D}. The instrument operates in vacuum at 73K, regulated to sub-mK level, with a temperature-controlled environment \citep{2018SPIE10702E..62C}. It has precise light injection devices \citep{2012SPIE.8446E..2EP} and splits light into two orthogonal polarization states using ZnSe quarter-wave rhombs and a Wollaston prism located in its Cassegrain unit. It uses two science fibers (A and B) and one calibration fiber (C) \citep{2018SPIE10702E..5RM}. The detector is a 15-micron science-grade HAWAII 4RG\textsuperscript{TM} (H4RG) from Teledyne Systems used in its up-the-ramp readout mode \citep{2018SPIE10709E..1PA}.

\subsection{SPIRou data reduction with \texttt{APERO}}

The spectra of GJ\,251 were reduced with the 0.7.275 version of the SPIRou data reduction system (DRS) called \texttt{APERO} (A PipelinE to Reduce Observations) which is detailed in \citet{2022ascl.soft11019C}. The \texttt{APERO} pipeline processes science frames to produce 2D and 1D spectra from the two science channels and the calibration channel. The wavelength calibration uses a combination of exposures from a UNe hollow cathode lamp and a Fabry-Pérot etalon \citep{2021A&A...648A..48H}. Barycentric Earth radial velocity (BERV) is calculated to shift the wavelengths to the barycentric frame of the solar system using \texttt{barycorrpy} \citep{2018RNAAS...2....4K}. Finally, the pipeline incorporates a correction for telluric absorptions and night-sky emission, which will be detailed in an upcoming publication (Artigau in prep.).

\subsection{LBL algorithm}

RVs of GJ\,251 were obtained from the telluric-corrected spectra using the line-by-line (LBL) method described in the paper by \citet{2022AJ....164...84A}. This method, which is based on the framework developed in \citet{2001A&A...374..733B}, has been proposed by \citet{Dumusque_2018} and further explored by \citet{Cretignier_2020} in the context of optical RV observations with HARPS. The LBL calculates Doppler shifts for individual spectral lines. To achieve this, a template with low noise is required, as the RVs are inferred from the comparison between the residuals of the observed spectrum and the template, as well as the derivative of the template. In practice, a high signal-to-noise ratio (SNR) combined spectrum is used as a template for a given target, to ensure that any remaining noise is small compared to an individual spectrum. The LBL method has been compared to other methods such as the cross-correlation function and template matching in \citet{2022A&A...660A..86M} and \citet{2022AJ....164...84A}.

\par 

For each spectrum, the LBL algorithm combines the velocities of individual lines (27702 lines in the case of GJ\,251) into a single RV measurement. This is achieved using a mixture model, which assumes that the velocities of individual lines are distributed according to a Gaussian distribution centered on the mean velocity, or another distribution of high-sigma outliers with multiple origins (e.g., bad pixels, cosmic rays, telluric residuals). The RV measurements are grouped by night by calculating the inverse-variance weighted mean of all observations for each night, resulting in a reduction from 769 to 177 data points. To address the variation in velocities of the day-to-day drift, a correction is applied using simultaneous Fabry-Pérot measurements. Furthermore, to take into account any long-term variation, a Gaussian Process (GP) regression is computed using the most observed stars in the SLS in order to estimate the zero point in the time domain.  This procedure is similar to the one conducted for SPIRou data in \citet{2022AJ....164...96C} or for SOPHIE \citep{2015A&A...581A..38C}, HARPS \citep{2020A&A...636A..74T} and HIRES \citep{2019MNRAS.484L...8T} spectrographs. In our case, we employed the \texttt{celerite} package \citep{celerite} to fit a GP model using a Matérn $3/2$ kernel. The estimation of the hyper-parameters was achieved by minimizing the likelihood function. Studies on common trends in SPIRou data are being conducted in parallel, and will address the differences between GPs and more commonly used tools such as a running median or nightly offsets in forthcoming work by the SPIRou team.

\par

\begin{figure}[!h]
    \centering
    \includegraphics[width=\linewidth]{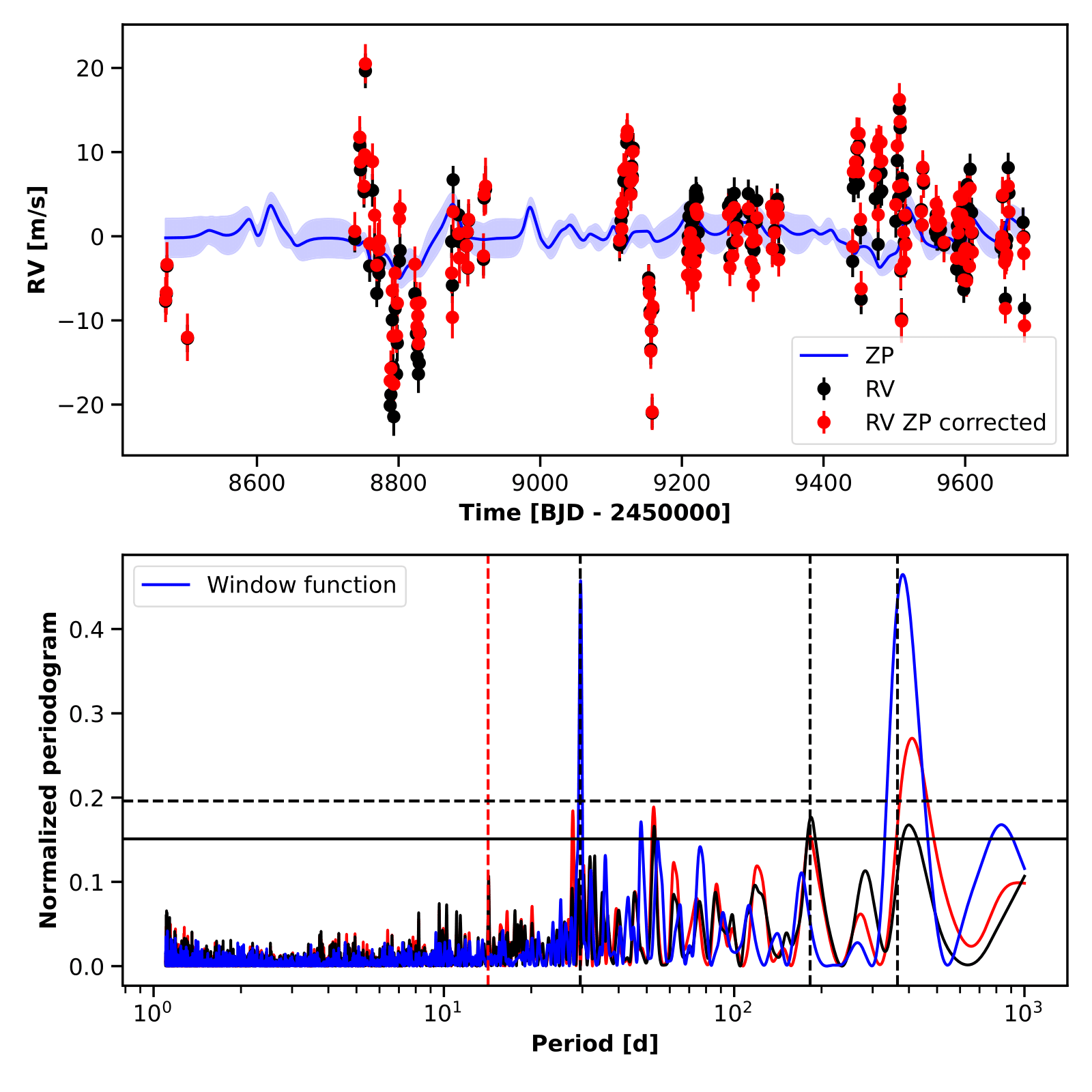}
    \caption{The top panel shows the data with and without zero point correction, indicated by black and red points respectively. The zero point offset is shown in blue. The bottom panel shows the corresponding periodograms for the time-series in the same color scheme, and the blue line represents the window function. The significance levels of FAP = $10^{-3}$ and FAP = $10^{-5}$ are indicated by the solid and dashed horizontal lines, respectively. The 1-year period, the 180 days period, and the synodic (29.5 days) period of the Moon are represented by vertical black dashed lines, and the period of the claimed planet in \citet{2020A&A...643A.112S} is in red.}
    \label{fig:zero_point}
\end{figure}

The final outcome of the data reduction is presented in Figure \ref{fig:zero_point}, wherein the red data points represent the results obtained after applying zero point correction and the black data points represent the results without such correction. To identify periodicities in the time-series, a generalized Lomb-Scargle (GLS) periodogram was used \citep{Zechmeister_2009}, with a grid of 100,000 points ranging from 1.1 days to 1,000 days, using the \texttt{astropy} package \citep{2018zndo...4080996T}. The false alarm probability (FAP) of the maximum peak was calculated using the analytical formula in \citet{2008MNRAS.385.1279B}.

\par 

As is illustrated, even with the careful processing steps in \texttt{APERO}, the LBL algorithm produces time series with spurious peaks at a period of 6 months and 1 year. However, the planet as reported in \citet{2020A&A...643A.112S} is not detected. Notably, the zero-point correction does not alleviate these systematic effects but rather amplifies them. As this 180-day peak is not present in the window function, it can be inferred that it does not arise from the observation strategy but from an unclear systematic effect. These signals may be attributed to a defective telluric correction; however, the underlying cause is poorly understood and a practical, physics-based correction is currently unavailable. Therefore, GJ\,251 serves as an appropriate example of how our method can eliminate spurious signals from RV data.

\section{The \texttt{Wapiti} method}\label{sec:method}

Our approach applies a principal component analysis to the per-line RV time-series produced by the LBL algorithm. Since these datasets may be noisy and contain missing data, we use the weighted principal component analysis (wPCA) method described in \citet{2015MNRAS.446.3545D}, implemented in the \texttt{wpca} module\footnote{https://github.com/jakevdp/wpca} and each observation is weighted by the inverse of its uncertainty squared. The objective of the \texttt{Wapiti} method is to eliminate the sources of systematic error. This is achieved by performing a wPCA reconstruction of the original per-line RV time-series and subtracting its reconstructions from it. The rationale behind this approach is that, in a time-series dominated by spurious signals of unknown origin, selecting appropriate components can serve as a proxy for correcting the effects responsible for the resulting systematic errors.

\par

To determine the appropriate number of principal vectors, we suggest the following approach. First, we discard all individual line RV time-series that have more than 50\% of missing data, reducing the number of lines from 27\,201 to 17\,923. Such missing data may occur for a line when the RV value deviates by more than 5 sigma from the standard deviation in relation to the line's RMS \citep{2022AJ....164...84A}. Subsequently, we bin each individual per-line RV time-series by night, by computing the inverse-variance weighted mean of all observations for each night. Afterward, we night-bin in a similar fashion the Fabry-Pérot drift also computed with the LBL and remove it from each individual per-lines RV time-series. Finally, we standardize the time-series
using their inverse-variance weighted mean and variance.

\subsection{Reducing the number of components through a permutation test}

\begin{figure}[!h]
    \centering
    \includegraphics[width=\linewidth]{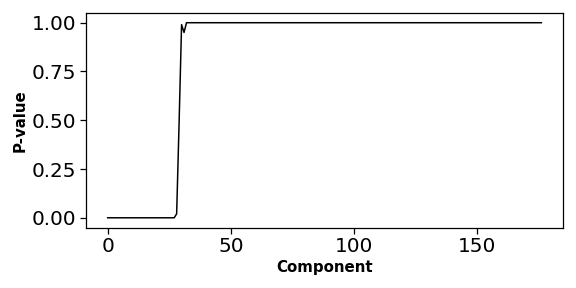}
    \caption{The figure shows the p-values of each component as determined by the permutation test. Only the components with a p-value below a certain threshold ($\le 10^{-5}$) are considered to have a significant contribution, rather than being just random noise.}
    \label{fig:p_value}
\end{figure}

The selection of the appropriate number of components for a wPCA is a critical aspect. Too few components may fail to effectively remove undesired contamination, while an excessive number of components may lead to over-fitting and an increase in the levels of random noise. To address this issue, a method is needed to determine which wPCA components are relevant and necessary for the analysis. In our work, we first use a permutation test \citep{permutation_test} which involves independently shuffling the columns of the dataset, and then conducting a wPCA on the resulting dataset. We repeat this process 100 times and compare the explained variance of the original dataset with the permuted versions. If a principal component is relevant, we expect that the permuted explained variances of that component will not be greater than the original explained variance. We can identify the relevant components by examining the p-values and selecting those with low p-values ($\le 10^{-5}$). The result is that only the first 26 components can not be explained by noise as it can be seen in Figure \ref{fig:p_value}.

\subsection{Leave-p-out cross-validation}

To evaluate the significance of the number of components retained after the permutation test, we conduct a leave-p-out cross-validation. This approach has been used by \citet{2022A&A...659A..68C} to select relevant components, albeit with variations in the methodology. In this study, we generate 100 subsets of the per-lines RV dataset, each containing 80\% of the lines randomly selected. The wPCA is then recalculated for each of these N datasets to assess the robustness of the 26 remaining components.

\par 

The underlying idea involves evaluating the robustness of the eigenvector basis, also known as principal vectors, by examining their stability under the removal of a few lines. Should a principal vector exhibit excessive sensitivity to the lines in use, it is deemed superfluous and may be discarded. We employ the Pearson coefficient of correlation as the evaluation metric, for the reasons stated in \citet{2022A&A...659A..68C} being that it is computationally efficient, bounded, and insensitive to the amplitudes of vectors. Specifically, we compute the Pearson coefficient between the n$^{\rm th}$ principal vector obtained from the wPCA using the full dataset and the n$^{\rm th}$ principal vector obtained from each of the N subsets. The absolute value of the coefficient is taken, as the wPCA vectors may change sign. The median Pearson coefficient value is calculated for each component number, and only those with a value above an arbitrary threshold of 95\% are retained, as in \citet{2022A&A...659A..68C}.

\par

\begin{figure}[!h]
    \centering
    \includegraphics[width=\linewidth]{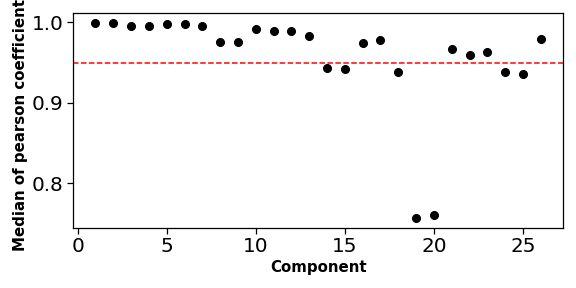}
    \caption{Mean Pearson coefficients as a function of the components. Those with a value below 95\% are discarded, leaving the number of components to 13.}
    \label{fig:pearson}
\end{figure}

As depicted in Figure \ref{fig:pearson}, it is observed that the value first goes below the threshold after the 13$^{\rm th}$ component. This indicates that the principal vectors after the 3$^{\rm th}$ one are not correlated with their corresponding vectors in the original dataset, implying that these eigenvectors are reliant on the lines used. As a result, they are disregarded, leaving us with 13 components for the remainder of the \texttt{Wapiti} method. In future research, efforts will be conducted to study how the number of components obtained through the application of the two methods, namely the permutation test and the leave-p-out cross-validation, varies with the target being studied.

\subsection{Removing outliers}

Per-line RVs have the potential to be affected by short-term fluctuations that are not correlated with the most significant principal vectors. These outliers can adversely affect the ability to identify systematics and reduce the accuracy of the reconstruction. \citet{2021MNRAS.505.1699C} applied  a singular value decomposition to the shape of the CCF and used a one-dimensional rejection mask in the time domain to remove outliers. For each principal component, they classified observations as "bad" if their absolute deviations are farther from the median value than a specified number $k_{mad}$ of the median absolute deviations (MAD), and "good" otherwise. If even one epoch in a principal component is an extreme outlier, it is likely that the entire observation is contaminated. In this study the authors determined that a $k_{mad}$ value of 6 yielded satisfactory results at the cost of rejecting 33 data points out of 886.

\par

In our work, we consider the per-line RV time-series and apply a similar method to the principal vectors. However, selecting an appropriate value for $k_{mad}$ is not a straightforward task. Although one could potentially choose a value such as $k_{mad}=15$ to correspond to a $10\sigma$ rejection, this would seem arbitrary. Furthermore, such an approach would not align with the aim of our method, which is to be able to operate independently on stars without requiring an intervention to select the appropriate value of $k_{mad}$ for each target. For this reason, we have chosen to employ a different strategy. We compute the ratio of the absolute deviation ($AD$) of the $n^{\rm nth}$ principal vector $V_n\left(t\right)$ to its $MAD$, and assign the maximum value of this ratio to each epoch across all components (we recall that for a timeseries $X$ and naming its median $\Bar{X}$ we have $AD\left(X\right) = |X - \Bar{X}|$ and $MAD = \texttt{median} \left(AD\left(X\right)\right)$). We use the letter $D$ to denote this value, which measures the degree of anomaly in an observation:

\begin{equation}
    D\left(t\right) = \max_{n} \frac{AD\left(V_n\left(t\right)\right)}{MAD\left(V_n\left(t\right)\right)}.
\end{equation}

Consequently, we reorganize the epochs in a descending order based on the value of $D$, resulting in a list that ranks the degree of anomaly of each observation. One could then choose to discard the first $N$ elements from this list. To identify the optimal value of $N$, we gradually remove them one by one and determine the number that maximizes the significance of the signal in the corrected RV once the \texttt{Wapiti} method has been applied. We found that removing the first 7 outliers produced the best outcomes, reducing the time series from 178 to 171 data points. It is acknowledged that readers may perceive this method to be specific to individual stars. However, we contend that this is a justifiable approach for our selected target GJ\,251, as the signal remains consistent regardless of the number of rejected outliers. Moreover, the main benefit of this approach is that it provides a framework to assess the level of anomaly for any target at a given epoch. For further information about this process, please refer to Appendix \ref{determiningkmad}.

\par 

\begin{figure*}[!h]
    \centering
    \includegraphics[width=\linewidth]{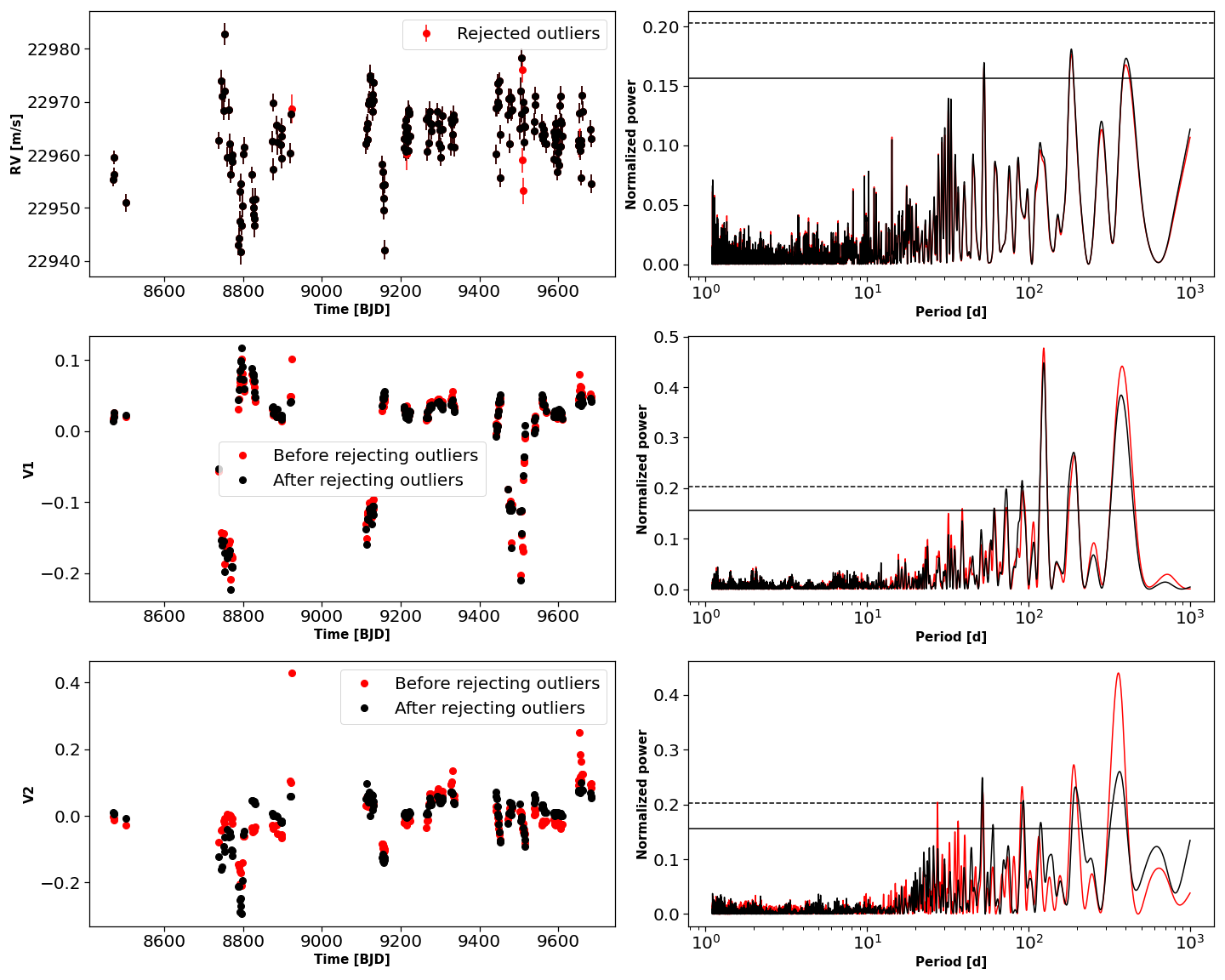}
    \caption{Left panels show the RV time-series and and the first two principal vectors. Their respective periodograms are on the right. The significance levels of FAP = $10^{-3}$ and FAP = $10^{-5}$ are indicated by the solid and dashed horizontal lines, respectively.}
    \label{fig:outliers_removal}
\end{figure*}

 The top-left panel of Figure \ref{fig:outliers_removal} displays the RV time-series with flagged outliers highlighted in red. The periodogram of the RV time-series after the removal of outliers is displayed on the right. Furthermore, the first two principal vectors are presented on the left, with the previous principal vectors using all available data points depicted in red, and the new principal vectors obtained from applying the wPCA on the remaining data points shown in black. The periodograms of the new principal vectors are displayed on the right. From the periodograms, it can be observed that the presence of a 1-year period and a 180-day period is evident in the two vectors, indicating that the wPCA method identified signals within the per-line RV dataset that may be contributing to the resulting systematic errors. We furthermore observed that the principal vectors did not exhibit any signs of the known 14.2\,d planet, indicating that our approach did not inadvertently capture the Keplerian signal in our data. It can also be noted that a prominent signal with a period of 120 days appears in the first vector, while not present in the RV time-series. This period could correspond to a harmonic of the 1 year signal, however we also note that it is close to the reported star' s rotation period of $P_{rot} = 124 \pm 5 $ days in \citet{2020A&A...643A.112S}. We attributed this signal to stellar activity for reasons that will be more detailed in subsequent sections.

\section{Results}\label{sec:results}

\subsection{The \texttt{Wapiti} corrected time-series and its analysis}

\begin{figure}[!h]
    \centering
    \includegraphics[width=\linewidth]{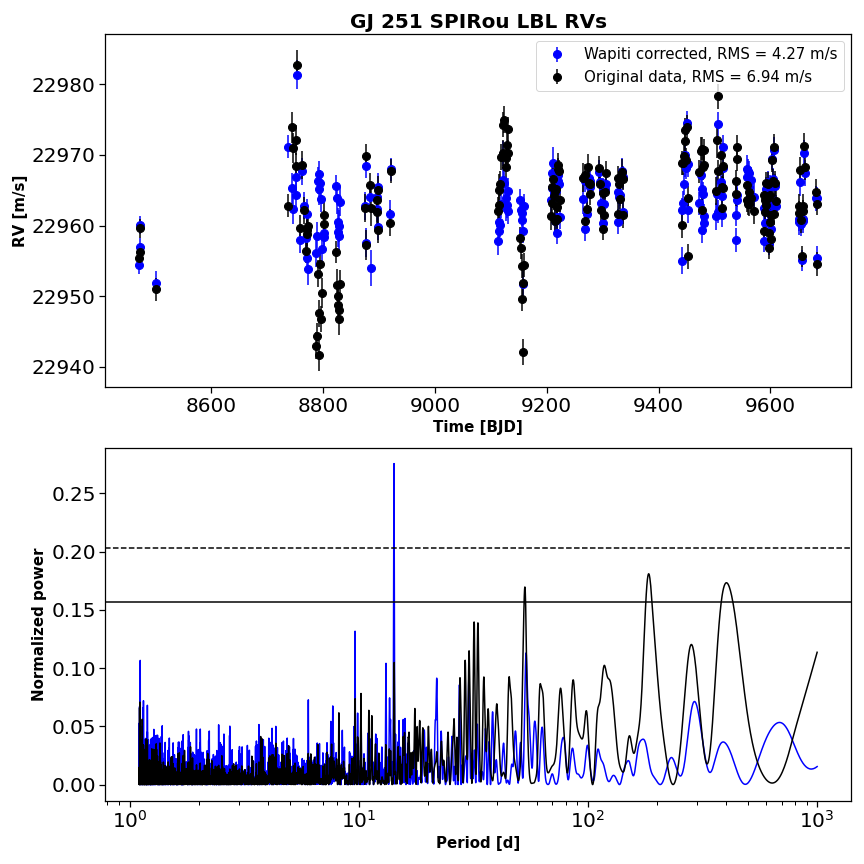}
    \caption{The effectiveness of the \texttt{Wapiti} correction method on the RV time-series. The top panel of the Figure shows the original data in black and the time-series from the \texttt{Wapiti} correction in blue. The bottom panel shows the corresponding periodograms using the same color code, it can be seen that the correction has successfully eliminated the 180d and 365d systematic peaks. The reported 14.2\,d period planet is now detected at high significance in the blue periodogram with a FAP level of $4.01 \times 10^{-9}$.}
    \label{fig:wapiti_corrected}
\end{figure}

The method outlined in the previous section has allowed us to determine the most appropriate number of components for our dataset on GJ\,251. We subsequently subtracted the reconstruction of the original time-series to itself to generate a new time-series, which we refer to as the \texttt{Wapiti} corrected time-series. We evaluated the effectiveness of the correction by analyzing the resulting periodogram as illustrated in Figure \ref{fig:wapiti_corrected}. In this Figure, the time-series resulting from the correction is depicted in blue and the original data are represented in black. The same color scheme is used when plotting the periodograms below the time series.

\par 

As desired, the \texttt{Wapiti} correction effectively eliminated the 180\,d and 365\,d peaks that we attributed to systematics present in our data since those periods are no longer statistically significant and the planet previously reported by \citet{2020A&A...643A.112S} is now discernible in our data, making it the first detection of GJ\,251b with near-IR velocities. More precisely, the most prominent peak went from a 184d period peak with a FAP level of $9.15 \times 10^{-5}$ to a 14.2\,d peak with a FAP level of $4.01 \times 10^{-9}$ after correction. In Appendix \ref{impact_component}, we explore the impact of the number of components used for the wPCA on this detection.

\par

\begin{figure}[!h]
    \centering
    \includegraphics[width=\linewidth]{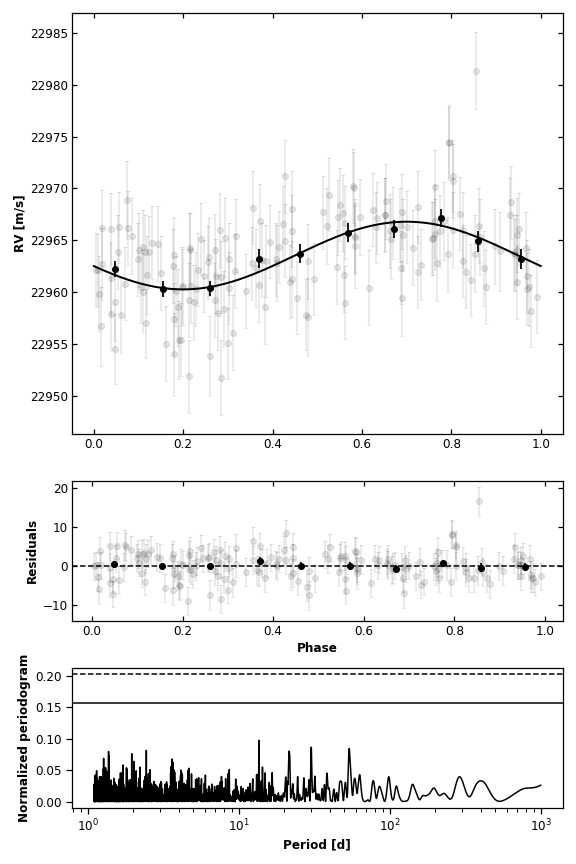}
    \caption{Results of the RV analysis using the \texttt{Wapiti} corrected time-series. The top panel illustrates the phase-folding of the planet with the best-fitted 14.2-day period, represented by a solid black curve, and the black circles depict the binned RV measurements. The middle panel displays the residuals after the model has been fitted, and the bottom panel presents the resulting periodogram, which indicates the absence of any remaining significant signals within the data.}
    \label{fig:rv_analysis}
\end{figure}

To analyze the corrected RV time-series obtained from the
\texttt{Wapiti} algorithm, a model $\mu$ consisting of a Keplerian orbit with a period of 14.2 days and an offset was employed, using the radvel package \citep{2018PASP..130d4504F} on the corrected data However, caution must be exercised when fitting this model, as the reconstruction process may degrade it. The reason being that although there were no signs of the Keplerian signal found in the principal vectors, it is important to note that they may not be orthogonal to it. Therefore, it is possible that when subtracting the wPCA projection from the time-series, the projection of the Keplerian signal could still have a non-zero value, posing a potential issue when trying to correctly characterize the signal. Therefore, it could be necessary to take into account the filtering effect of the wPCA reconstruction when fitting the model. This can be achieved with minimal computational cost by using the wPCA to reconstruct the desired model and then fitting the difference between the original model $\mu$ and its wPCA reconstruction $\hat{\mu}$ to the corrected RV time-series obtained from Wapiti. Concretely, denoting $R$ the wPCA reconstruction we have $\hat{\mu} = R\left(\mu\right)$ and we suggest to fit $\mu - \hat{\mu}$ to the corrected data instead of simply $\mu$. This process bears resemblance to that described in \citet{2020MNRAS.493.2215G}, wherein they used this method to take into account the filtering impact of the SysRem algorithm \citep{2005MNRAS.356.1466T} on their model. The periodogram of the residuals in \ref{fig:rv_analysis} does not show any signs of one of the planets reported in \citet{2017AJ....153..208B}.
\par

To obtain estimates of the uncertainties on the free parameters, we employed a Bayesian Markov Chain Monte Carlo (MCMC) framework to sample the posterior distribution, with 100\,000 samples, 10\,000 burn-in samples, and 100 walkers using the \texttt{emcee} module \citep{2013ascl.soft03002F}. The orbital and derived planetary parameters are shown in Table \ref{gl251_parameters} and compared to those obtained in \citet{2020A&A...643A.112S}. In order to calculate the mass of the planet, $M_p \sin i$, we used the mass of the star and its uncertainty obtained from \cite{2022MNRAS.516.3802C} and listed in Table \ref{gl251_stellar_properties}. One may note that the eccentricity is poorly constrained from our RV data, prompting the question of whether a circular fit would yield a better result. To address this, we conducted a comparison of the maximum likelihood estimation (MLE) for both Keplerian and circular models, allowing us to compute their respective Bayesian Information Criterion (BIC; see \citet{1978AnSta...6..461S}). By taking the logarithm of the Bayes factor ($\log\ BF = \frac{BIC_{circ.} - BIC_{kep.}}{2}$), we were able to compare the two models. Our analysis strongly favored the Keplerian model ($\log\ BF = 15$), and as such, we decided not to perform a MCMC estimation of the orbital parameters using a circular model.

\begin{table*}[t]
\centering
\begin{tabular}{c  c  c  c}
\hline
& \texttt{Wapiti} & Priors & Stock et al. (2020)  \\ [0.5ex]
\hline
\noalign{\smallskip}
\multicolumn{4}{c}{\textbf{Orbital Parameters}} \\
\noalign{\smallskip}
\hline
\noalign{\smallskip}
Period (days) & $14.24^{+0.01}_{-0.01}$ & $\mathcal{U}\left(14,15\right)$ & $14.238^{+0.002}_{-0.002}$ \\
\noalign{\smallskip}
$t_0 - 2450000$ (BJD) & $8628^{+3}_{-3}$ & $\mathcal{U}\left(8622.01,8632.01\right)$ & $8626.69^{+0.34}_{-0.35}$ \\
\noalign{\smallskip}
$K$ (m\,s$^{-1}$) & $3.25^{+0.41}_{-0.40}$ & $\mathcal{U}\left(0,4\right)$ & $2.11^{+0.21}_{-0.20}$ \\
\noalign{\smallskip}
$\sqrt{e}\cos{\omega}$ & $-0.12^{+0.32}_{-0.23}$ & $\mathcal{U}\left(-1,1\right)$ & $0.06^{+0.21}_{-0.22}$ \\
\noalign{\smallskip}
$\sqrt{e}\sin{\omega}$ & $0.10^{+0.22}_{-0.22}$ & $\mathcal{U}\left(-1, 1\right)$ & $0.20^{+0.16}_{-0.22}$ \\
\noalign{\smallskip}
\hline
\noalign{\smallskip}
\multicolumn{4}{c}{\textbf{Derived Parameters}} \\
\noalign{\smallskip}
\hline
\noalign{\smallskip}
$M_p \sin i$ (M${\oplus}$) & $6.02^{+0.80}_{-0.78}$ & - &  $4.00^{+0.40}_{-0.40}$ \\ 
\noalign{\smallskip}
$\omega$ (degrees) & $60^{+86}_{-209}$ & - &  $78.8^{+47.6}_{-44.7}$ \\ 
\noalign{\smallskip}
$e$& $0.11^{+0.11}_{-0.07}$ & - & $0.10^{+0.09}_{-0.07}$ \\ 

\noalign{\smallskip}
\hline
\end{tabular}
\caption{Orbital and derived planetary parameters from the analysis of GJ\,251. The parameters derived from the \texttt{Wapiti} corrected RV data from SPIRou are compared to those from \citet{2020A&A...643A.112S} using CARMENES and HIRES spectrographs.}
\label{gl251_parameters}
\end{table*}

\subsection{Injection recovery test}

In order to assess the effectiveness of the \texttt{Wapiti} method in retrieving correct planetary parameters and to verify our claim that it is necessary to take into account the wPCA reconstruction of our model, we performed an injection recovery test. We first removed the \texttt{Wapiti} corrected RV time-series from each per-line RV time-series. We then injected a signal consisting of a Keplerian with the parameters corresponding to the best-fitted parameters of GJ\,251b from our analysis, except for the semi-amplitude $K_{injected}$ that we made vary from $2$ to $10$ m\,s$^{-1}$.

\par

We chose a range of semi-amplitude values that explore the low and intermediate regime where the planetary signal is either of lower or of comparable magnitude with the systematics present in our data. In our case, the systematics consist of a $180$\,d signal of $\sim 6$ m\,s$^{-1}$ amplitude. This range of semi-amplitude allowed us to assess the ability of the \texttt{Wapiti} method to recover the injected planetary signal.

\par 

For each injected signal, we employed a simple model comprising of only the injected Keplerian with varying semi-amplitude and an offset. Our aim was to quantitatively assess the effect of the \texttt{Wapiti} method on the estimated semi-amplitude $K_{computed}$. To accomplish this, we used a MCMC approach, which consisted of 100,000 steps, 10,000 burn-in samples, and 100 walkers. We set a uniform prior distribution between -50 and 50 m\,s$^{-1}$ over the semi-amplitude.

\par 

\begin{figure}[!h]
    \centering
    \includegraphics[width=\linewidth]{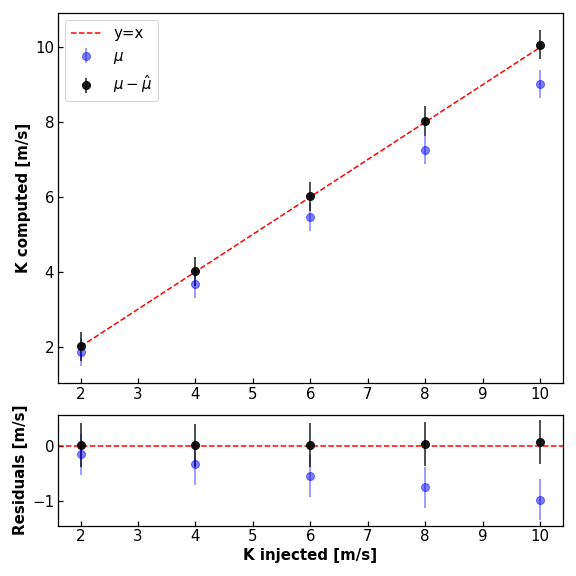}
    \caption{Top panel displays the retrieved semi-amplitude $K_{computed}$ versus the injected semi-amplitude $K_{injected}$. The bottom panel illustrates the residuals, demonstrating the successful recovery of the injected signal using the \texttt{Wapiti} method, with the inclusion of the wPCA reconstruction in the fitted model.}
    \label{fig:injection_recovery_test}
\end{figure}

In order to evaluate the importance of subtracting the wPCA reconstruction $\hat{\mu}$ from the model $\mu$, we conducted a comparative analysis by fitting both $\mu$ and $\mu - \hat{\mu}$ to the data and examining their impact on the estimated semi-amplitude. Our findings displayed in figure \ref{fig:injection_recovery_test} demonstrate that the injection recovery test is generally successful and the \texttt{Wapiti} method accurately recovers the correct semi-amplitude. However, as the semi-amplitude increases, the exclusion of the reconstruction $\hat{\mu}$ becomes increasingly significant. Fitting only $\mu$ results in an underestimation of the correct semi-amplitude.

\subsection{Stellar activity in GJ\,251}

It is noteworthy that contrary to the method employed by \citet{2020A&A...643A.112S}, we did not use a quasi-periodic GP to account for stellar activity in our analysis. In their paper, they reported a rotation period of $124 \pm 5$ days while \citet{2023A&A...672A..52F} detected a periodicity of $98.7_{-4.8}^{+11.5}$ days by analyzing the time-series of the longitudinal magnetic field ($B_\ell$), yet we did not observe such signals in our RV periodogram. Nonetheless, for completeness, we fitted two models consisting of one planet and one planet plus a quasi-periodic GP of the form

\begin{equation}
    k\left(\tau\right) = A \exp\left(- \frac{\tau^2}{2l^2} - \Gamma \sin^2\left(\frac{\pi}{P}\tau \right)\right).
    \label{gp_form}
\end{equation}

The GP was computed using \texttt{george} \citep{2015ITPAM..38..252A}. We calculated the Bayes factor to compare the two models and it was found that adding the quasi-periodic GP did not improve the fit ($\log\ BF = -5.5$), reinforcing our conclusion that stellar activity does not impact our RV data. Ultimately, we determined that it was not necessary to account for the influence of stellar activity when characterizing the planet orbiting GJ\,251.

\par 

\begin{figure*}[!h]
    \centering
    \includegraphics[width=\linewidth]{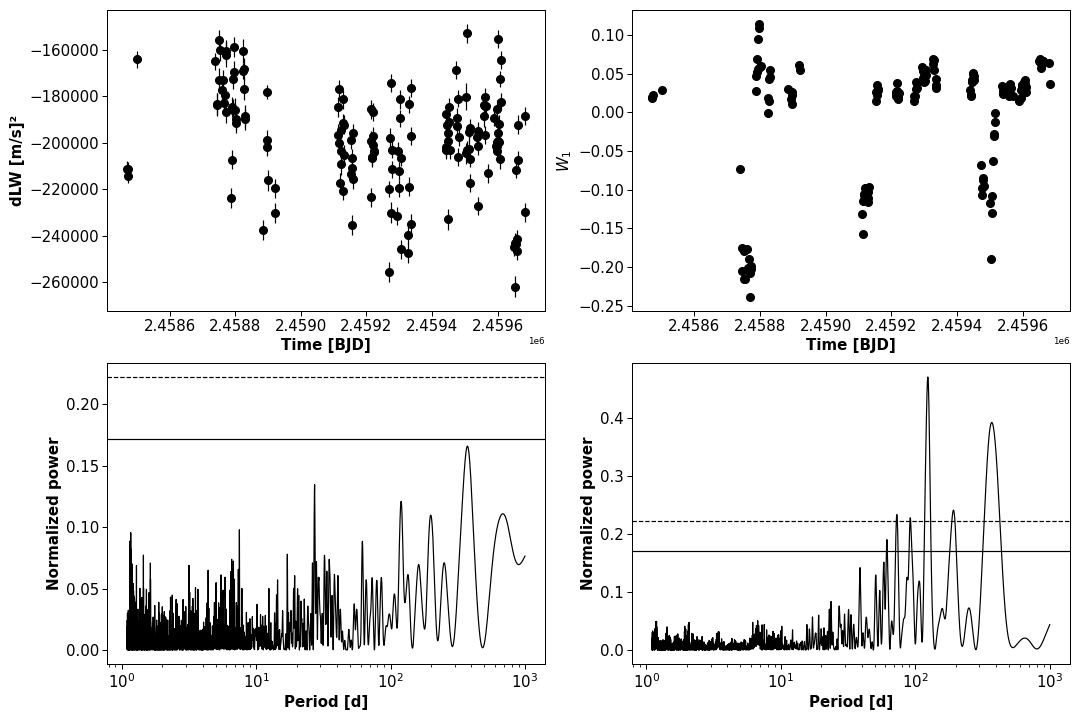}
    \caption{The  left panels display, at the top, the dLW time-series and its corresponding periodogram at the bottom, which shows no indication of periodicity in the data. On the right, the first principal component of the per-line dLW time-series obtained from the wPCA is displayed, which exhibits a pronounced 124-day period peak.}
    \label{fig:activity}
\end{figure*}

However, when performing a wPCA on the differential line width (dLW; introduced in \citet{2018A&A...609A..12Z}), which is another data product obtained from the LBL analysis that measures variations in the average width of line profiles relative to a template \citep{2022AJ....164...84A}, Cadieux et al. (in prep.) found that the first principal vectors can be sensitive to the presence of stellar activity. In our case, we observed that even though no periodicity can be detected in dLW, the first principal vector ($W_1$) presents a signal that is similar to the one reported in \citet{2020A&A...643A.112S} as displayed in Figure \ref{fig:activity}. We fitted a quasi-periodic GP in order to estimate the value of the periodicity in $W_1$ fitting a quasi-periodic kernel of the same form as in Equation \ref{gp_form} as well as a jitter term and a constant. We computed an MCMC using 10,000 steps, 1,000 burn-in samples, and 100 walkers which revealed a periodicity of $137^{+37}_{-16}$ days comparable within uncertainties to the one obtained in \citet{2020A&A...643A.112S}. This led us to conclude that $W_1$ could serve as a good proxy for stellar activity and the specifics of this analysis can be found in Appendix \ref{W1_analysis}. In addition, in Donati et al. (2023, submitted) it was found when studying AUMic that the first time derivative of $W_1$ correlated well with the RV which is not the case in our study (Pearson coefficient of -0.11). That said, a noteworthy result is the correlation between the first principal component of the wPCA applied to the RVs and $W_1$ with a Pearson coefficient of 0.96, suggesting that the secondary peak of the first principal vector at 120d may be related to stellar activity and not be a 1 year harmonic. This highlights that despite the lack of any clear signs of activity in both RV and dLW time-series, the presence of such activity can still be detected in their principal vectors. The results from both our data and \citet{2020A&A...643A.112S} demonstrate that the impact of stellar activity on RV data can manifest with a $\sim 2 \sigma$ difference to the periodicity reported in \citet{2023A&A...672A..52F} when studying $B_\ell$. This indicates that the effect of stellar activity on RV data may not always occur at the same period, for instance due to differential rotation, and must be evaluated with available indicators on a case-by-case basis.

\section{Discussion, summary, and prospects}\label{sec:ccl}

The \texttt{Wapiti} method successfully eliminated the spurious signals in the GJ\,251 data and detected the planetary signal. Although our study demonstrated the effectiveness of the method in accurately estimating planetary parameters, it is important to carefully consider the characteristics of the data. For instance, in situations where a high semi-amplitude planetary signal or trend is present, it would be recommended to first fit these signals and then apply the method to the residuals.

\par 

Furthermore, due to its data-driven nature, this method does not require a model for the origins of the corrected systematic errors. Nevertheless, the \texttt{Wapiti} method does not aim to replace more physically motivated approaches, such as the zero point correction (which in the instance of GJ\,251 is not sufficient), but rather to be a tool that works in conjunction with them. In fact, this method is quite helpful and convenient when investigating the sources of systematics, as outlined in the following discussion.

\subsection{A BERV effect in SPIRou LBL RVs of GJ\,251}

\begin{figure}[!h]
    \centering
    \includegraphics[width=\linewidth]{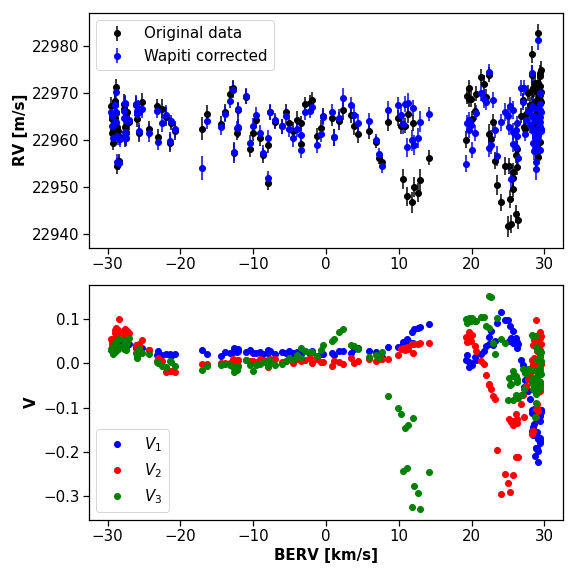}
    \caption{The top panel displays the original data's variability with respect to the BERV and how our method removed this dependency in the \texttt{Wapiti} corrected RV. The bottom panel demonstrates the behavior of the first three principal vectors with the BERV.}
    \label{fig:berv_effect}
\end{figure}

From the outset of our study, it was apparent that the spurious signals at 180 and 365 days in our dataset could be attributed to a residual telluric effect, as verified through visual examination of the data in the BERV space in Figure \ref{fig:berv_effect}. This analysis revealed the presence of a BERV effect, with significant fluctuations occurring in the original data at BERV values of approximately 10 and 25 km\,s$^{-1}$. It is worth noting that the largest fluctuation observed in our data corresponds to a velocity of the star that is close to zero in the observer's frame of reference due to the star's velocity of $22.98^{+0.22}_{-0.22}$ km\,$s^{-1}$.

\par

The principal vectors successfully captured the BERV dependency, as evidenced by an evident structure in the BERV space. For the sake of readability we only displayed the first three vectors in Figure \ref{fig:berv_effect}. Therefore, it can be concluded that the \texttt{Wapiti} method was able to correct for systematics in the original data by removing the BERV dependency through substracting the wPCA reconstruction. Although the origin of this BERV effect in the reduced data remains somewhat unclear, the next subsection describes how the \texttt{Wapiti} products can be used to investigate the source of systematics.

\subsection{Investigating the sources of systematics}

An approach to search for the underlying physical causes of systematic variations in our data would be to examine the scores obtained from the wPCA. These scores represent the transformed data in the new coordinate system established by the principal vectors. To quantify the influence of a specific principal component on a given line, one can examine its score values $U_n$. We propose a method that involves computing the value $\Tilde{U}_n = \frac{U_n - \Bar{U}_n}{\sigma \left(U_n\right)}$, where $\Bar{U}_n$ and $\sigma \left(U_n\right)$ denote the weighted mean and uncertainties of $U_n$, respectively. For each line $\lambda$, we determine the maximum absolute Z-score value of $\Tilde{U}_n\left(\lambda \right)$ over all the components. Recall that the Z-score of a value $x$ in a sample with mean $\mu$ and standard deviation $\sigma$ is given by $Z(x) = \frac{x-\mu}{\sigma}$. Hence, the metric we employ to quantify the extent to which a line is affected by the systematics is determined for each line $\lambda$ as follows:

\begin{equation}
    z\left(\lambda\right) = \max_{n} | Z\left(\Tilde{U}_n\left(\lambda \right)\right)|.
\end{equation}

\par

\begin{figure*}[!h]
    \centering
    \includegraphics[width=\linewidth]{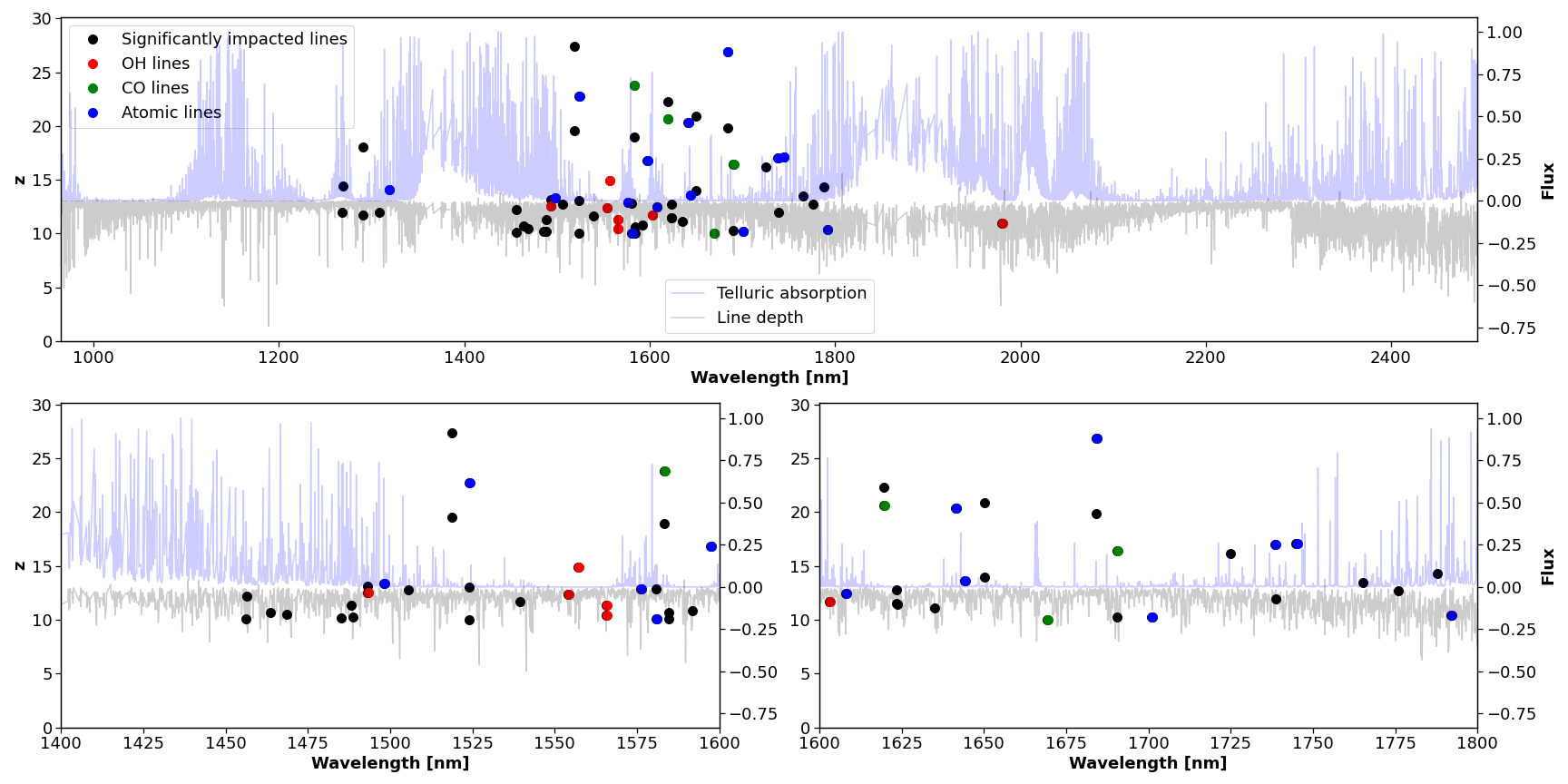}
    \caption{In the top-left panel, the lines that are significantly affected by the component are displayed in black. If these lines could be attributed to a molecular line, they are color-coded in red, while if they correspond to an atomic line, they are colored in blue. The telluric absorption computed with \texttt{TAPAS} is also displayed, as well as the template of GJ\,251 used by the LBL algorithm. The two lower panels provide zoomed-in views from 1400 to 1800\,nm.}
    \label{fig:score_first_component}
\end{figure*}

The higher the value of $z$, the more a line can be considered to be affected by at least one of the 13 components. In Figure \ref{fig:score_first_component}, we have constructed a plot that displays the lines with a $z$ value above an arbitrary threshold of $z_0 = 10$. This selection criterion resulted in a total of 63 lines being displayed on the plot. Additionally, we have overplotted in blue a synthetic telluric spectrum obtained using \texttt{TAPAS} \citep{2014hitr.confE...8B} available on the \texttt{APERO} GitHub repository \footnote{https://github.com/njcuk9999/apero-drs/tree/main/apero/data/spirou/telluric} as well as the template of GJ\,251 that was used to perform the LBL algorithm.

\par 

Figure \ref{fig:score_first_component} reveals that the lines most affected by the systematics are predominantly situated within the $H$ band (1500-1800\,nm), as well as in the immediate vicinity of the telluric regions surrounding this band. However, we note that 7 of them are located outside this shared region. To try to establish a connection between affected lines and stellar lines, we computed atomic and molecular lines with \texttt{VALD} \citep{2017ASPC..510..518P} for a star with a temperature of 3500\,K. The atomic mask consists of 2\,932 lines from 24 elements (H, Na, Mg, Al, Si, S, K, Ca, Sc, Ti, V, Cr, Mn, Fe, Co, Ni, Cu, Ga, Rb, Sr, Y, Zr, Ba, Lu). As for the molecular mask, it includes 4\,666 lines from 3 elements (OH, CO, CN). In Figure \ref{fig:score_first_component}, we employed a color-coding scheme to highlight any line from the two masks that met two criteria: first, it fell within the start and end points of an affected line, and second, its depth was above $0.7\%$ to avoid being considered as noise due to the SNR of our observations. The depth of the line being determined as the contrast of the line's background in comparison to the mean of the boundaries before and after it. Based on our analysis, we observed that 14 atomic lines (in blue), 8 OH lines (in red) and 9 CO lines (in green) could correspond to the affected lines, which accounts for 50\% of the total affected lines. Furthermore, we observed that 27 of those lines that have a potential association with the star are located in the $H$ band. For the remaining lines located in the telluric region, our findings suggest a possible complex interplay between telluric and stellar lines. Specifically, the overlapping of molecular or atomic lines with telluric lines may explain why both a 1-year signal and the star's rotational period can be distinguished in $V_1$.

\par

However, the origin of the 32 remaining lines with a $z$ value above $10$ is uncertain, some of them could be molecular lines as of yet unidentified but also of a completely different origin. While telluric or stellar activity could potentially be contributing factors, they may not fully explain the observed impact, suggesting the presence of other sources of spurious signals in the data. The complex interplay between different sources of systematics makes it challenging for the wPCA to distinguish them clearly, particularly since the wPCA is a linear method that can not differentiate between non-linear relationships between features in the data. Therefore, should the interplay between overlapping stellar lines and telluric lines be non-linear, we may encounter such an issue. Ultimately, the purpose of this discussion is not to conclusively explain the origin of the systematics, but rather to demonstrate how the results obtained from the \texttt{Wapiti} method can lead to a better understanding of them. In order to identify the exact cause of the systematics, it would be necessary to conduct an investigation of the scores associated with the principal vectors across multiple targets, which is beyond the scope of this study.

\par 

\begin{figure}[!h]
    \centering
    \includegraphics[width=\linewidth]{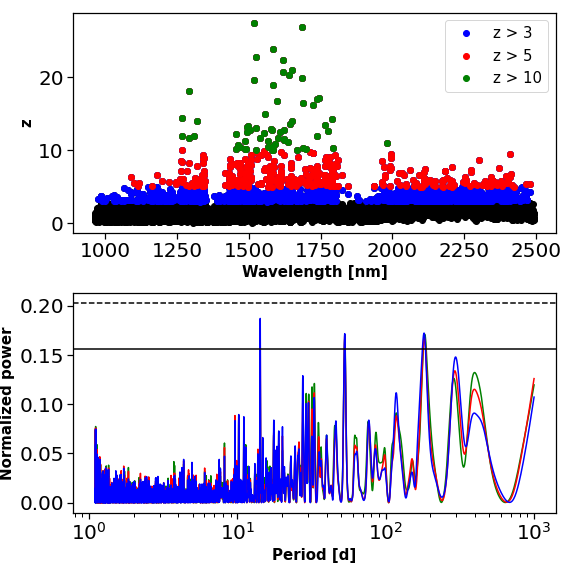}
    \caption{Top panel shows the $z$ metric of all lines. Bottom panel displays the periodogram of the resulting time-series in green, red and blue when respectively rejecting lines with a $z$ value above 10, 5 and 3.}
    \label{fig:lines_removal}
\end{figure}

One may wonder if those systematics only impact a few lines and what would happen if we were to simply remove those high impacted lines from the calculation when computing the final RV time-series. We tested this idea by removing lines for various threshold values of $z$. We computed the periodogram of the resulting time-series and we see in Figure \ref{fig:lines_removal} that even a very strict criterion of removing lines with $z \geq 3$ only very weakly enhanced the 14.2\,d period signal ($FAP = 4.9 \times 10^{-5}$) while not effectively removing the systematics from the RV data, at the cost of removing 1\,939 lines from the computation. These results suggest that simply removing the most impacted lines does not eliminate systematics from the RV time-series. Instead, the wPCA reconstruction performed by the \texttt{Wapiti} method is necessary to successfully eliminate the systematics

\par 

Further research could explore the potential of variational autoencoders (VAE, \citet{2013arXiv1312.6114K}) to reconstruct the per-line RV time-series, which may be able to capture non-linear relationships between the systematic errors. Previous research has explored the use of neural networks in the examination of RV time-series by incorporating both RV and stellar activity indicators as input \citep{2023A&A...672A.118P, 2023MNRAS.519.5439C}. However, the present study concentrates on the performance of the \texttt{Wapiti} method and demonstrates that a wPCA, more straightforward, easier to follow, and more efficient computationally, already produces promising outcomes. Additional investigation would be necessary to assess the advantages of applying a VAE to per-line RV time-series.

\subsection{Conclusion and future work}

In this paper, we have presented \texttt{Wapiti}, a new data-driven algorithm for correcting systematic errors in RV data. The algorithm uses a dataset of per-line RV time-series generated by the LBL algorithm and employs a weighted principal component analysis to subtract the wPCA reconstruction from the original data. This process creates a new RV time-series free of systematic errors. To determine the appropriate principal vectors, the algorithm incorporates a permutation test followed by leave-p-out cross-validation algorithm. Once the relevant number of components have been identified an outlier rejection step follows and the wPCA is applied to the remaining data points. The wPCA reconstruction is consequently removed from original RV time-series, allowing for the derivation of a new \texttt{Wapiti} corrected RV time-series.

\par 

The application of \texttt{Wapiti} to GJ\,251 allows to successfully free the RV data from spurious signals at 6 months and 1 year while allowing the detection of a known temperate super-Earth with a period of 14.2 days. An examination of the sources of corrected systematics, as revealed by the $z$ metric, suggests that the method is able to address various sources of spurious signals in the data, including but not limited to stellar activity, telluric lines, and other unidentified sources. This highlights the potential for SPIRou to detect exoplanets, even in the presence of known or unknown sources of systematics. Furthermore, the algorithm has the potential to enhance the precision and accuracy of RV data from future exoplanet detection surveys, thereby advancing our understanding of the diversity and frequency of exoplanetary systems. Finally, this work provided further evidence to support the idea that the first principal vector $W_1$ derived from the wPCA of the dLW time-series can be employed as a reliable activity indicator to detect the stellar rotation period in our data.

\par 

While the \texttt{Wapiti} algorithm has shown encouraging results, there are still some remaining questions and challenges that need to be addressed. Future research should aim to evaluate the algorithm on a broader range of RV data and exoplanetary systems and better understand the sensitivity of planetary parameters to the method. A specific direction for future work will be to apply the \texttt{Wapiti} algorithm to targets of the SLS-Planet Search program that exhibit systematic signals of unknown origin at similar periods.

\begin{acknowledgements}

This project received funding from the European Research Council under the H2020 research \& innovation program (grant \#740651 NewWorlds).
\par
E.M. acknowledges funding from FAPEMIG under project number APQ-02493-22 and research productivity grant number 309829/2022-4 awarded by the CNPq, Brazil.
\par
 J.H.C.M. is supported in the form of a work contract funded by Fundação para a Ciência e Tecnologia (FCT) with the reference DL 57/2016/CP1364/CT0007; and also supported from FCT through national funds and by FEDER-Fundo Europeu de Desenvolvimento Regional through COMPETE2020-Programa Operacional Competitividade e Internacionalização for these grants UIDB/04434/2020 \& UIDP/04434/2020, PTDC/FIS-AST/32113/2017 \& POCI-01-0145-FEDER-032113, PTDC/FIS-AST/28953/2017 \& POCI-01-0145-FEDER-028953, PTDC/FIS-AST/29942/2017. 
\par 
 This work has been carried out within the framework of the NCCR PlanetS supported by the Swiss National Science Foundation under grants 51NF40\_182901 and 51NF40\_205606.
\par
This project has received funding from the European Research Council (ERC) under the European Union's Horizon 2020 research and innovation programme (grant agreement SCORE No 851555)
\par
We acknowledge funding from Agence Nationale pour la Recherche (ANR, project ANR-18-CE31-0019 SPlaSH)
\par
X.D. and A.C. acknoweldge funding by the French National Research Agency in the framework of the Investissements d’Avenir program (ANR-15-IDEX-02), through the funding of the “Origin of Life” project of the Université Grenoble Alpes.
\par
BK ackowledges funding from the European Research Council (ERC) under the H2020 research \& innovation programme (grant agreements \#865624 GPRV).
\par
This research made use of the following software tools: \texttt{NumPy}; a fundamental package for scientific computing with Python. \texttt{pandas}; a library providing easy-to-use data structures and data analysis tools. \texttt{Astropy}; a community-developed core Python package for Astronomy. \texttt{SciPy}; a library for scientific computing and technical computing. \texttt{RadVel}; a python package for modeling radial velocity data. \texttt{matplotlib}; a plotting library for the Python programming language. \texttt{tqdm}; a library for creating progress bars in the command line. \texttt{seaborn}; a data visualization library based on matplotlib. \texttt{wpca}; a python package for weighted principal component analysis.  \texttt{emcee}; a python package for MCMC sampling. \texttt{PyAstronomy}; a collection of astronomical tools for Python. \texttt{celerite}; a fast Gaussian process library in Python.

\end{acknowledgements}

\bibliographystyle{aa}
\bibliography{ref.bib}{}

\begin{thebibliography}{66}
\expandafter\ifx\csname natexlab\endcsname\relax\def\natexlab#1{#1}\fi

\bibitem[{{Ambikasaran} {et~al.}(2015){Ambikasaran}, {Foreman-Mackey},
  {Greengard}, {Hogg}, \& {O'Neil}}]{2015ITPAM..38..252A}
{Ambikasaran}, S., {Foreman-Mackey}, D., {Greengard}, L., {Hogg}, D.~W., \&
  {O'Neil}, M. 2015, IEEE Transactions on Pattern Analysis and Machine
  Intelligence, 38, 252

\bibitem[{{Anglada-Escud{\'e}} {et~al.}(2016){Anglada-Escud{\'e}}, {Amado},
  {Barnes}, {Berdi{\~n}as}, {Butler}, {Coleman}, {de La Cueva}, {Dreizler},
  {Endl}, {Giesers}, {Jeffers}, {Jenkins}, {Jones}, {Kiraga}, {K{\"u}rster},
  {L{\'o}pez-Gonz{\'a}lez}, {Marvin}, {Morales}, {Morin}, {Nelson}, {Ortiz},
  {Ofir}, {Paardekooper}, {Reiners}, {Rodr{\'\i}guez},
  {Rodr{\'\i}guez-L{\'o}pez}, {Sarmiento}, {Strachan}, {Tsapras}, {Tuomi}, \&
  {Zechmeister}}]{2016Natur.536..437A}
{Anglada-Escud{\'e}}, G., {Amado}, P.~J., {Barnes}, J., {et~al.} 2016, \nat,
  536, 437

\bibitem[{{Artigau} {et~al.}(2014){Artigau}, {Astudillo-Defru}, {Delfosse},
  {Bouchy}, {Bonfils}, {Lovis}, {Pepe}, {Moutou}, {Donati}, {Doyon}, \&
  {Malo}}]{2014SPIE.9149E..05A}
{Artigau}, {\'E}., {Astudillo-Defru}, N., {Delfosse}, X., {et~al.} 2014, in
  Society of Photo-Optical Instrumentation Engineers (SPIE) Conference Series,
  Vol. 9149, Observatory Operations: Strategies, Processes, and Systems V, ed.
  A.~B. {Peck}, C.~R. {Benn}, \& R.~L. {Seaman}, 914905

\bibitem[{{Artigau} {et~al.}(2022){Artigau}, {Cadieux}, {Cook}, {Doyon},
  {Vandal}, {Donati}, {Moutou}, {Delfosse}, {Fouqu{\'e}}, {Martioli}, {Bouchy},
  {Parsons}, {Carmona}, {Dumusque}, {Astudillo-Defru}, {Bonfils}, \&
  {Mignon}}]{2022AJ....164...84A}
{Artigau}, {\'E}., {Cadieux}, C., {Cook}, N.~J., {et~al.} 2022, \aj, 164, 84

\bibitem[{{Artigau} {et~al.}(2018){Artigau}, {Saint-Antoine}, {L{\'e}vesque},
  {Vall{\'e}e}, {Doyon}, {Hernandez}, \& {Moutou}}]{2018SPIE10709E..1PA}
{Artigau}, {\'E}., {Saint-Antoine}, J., {L{\'e}vesque}, P.-L., {et~al.} 2018,
  in Society of Photo-Optical Instrumentation Engineers (SPIE) Conference
  Series, Vol. 10709, High Energy, Optical, and Infrared Detectors for
  Astronomy VIII, ed. A.~D. {Holland} \& J.~{Beletic}, 107091P

\bibitem[{{Baluev}(2008)}]{2008MNRAS.385.1279B}
{Baluev}, R.~V. 2008, \mnras, 385, 1279

\bibitem[{{Becerril} {et~al.}(2017){Becerril}, {Mirabet}, {Lizon}, {Calvo},
  {Abril}, {C{\'a}rdenas}, {Ferro}, {Morales}, {P{\'e}rez}, {Ram{\'o}n},
  {S{\'a}nchez-Carrasco}, {Quirrenbach}, {Amado}, {Ribas}, {Reiners},
  {Caballero}, {Seifert}, \& {Herranz}}]{2017MS&E..278a2191B}
{Becerril}, S., {Mirabet}, E., {Lizon}, J.~L., {et~al.} 2017, in Materials
  Science and Engineering Conference Series, Vol. 278, Materials Science and
  Engineering Conference Series, 012191

\bibitem[{{Bertaux} {et~al.}(2014){Bertaux}, {Lallement}, {Ferron}, \&
  {Boonne}}]{2014hitr.confE...8B}
{Bertaux}, J.-L., {Lallement}, R., {Ferron}, S., \& {Boonne}, C. 2014, in 13th
  International HITRAN Conference, 8

\bibitem[{{Bonfils} {et~al.}(2013){Bonfils}, {Delfosse}, {Udry}, {Forveille},
  {Mayor}, {Perrier}, {Bouchy}, {Gillon}, {Lovis}, {Pepe}, {Queloz}, {Santos},
  {S{\'e}gransan}, \& {Bertaux}}]{2013A&A...549A.109B}
{Bonfils}, X., {Delfosse}, X., {Udry}, S., {et~al.} 2013, \aap, 549, A109

\bibitem[{{Bonfils} {et~al.}(2007){Bonfils}, {Mayor}, {Delfosse}, {Forveille},
  {Gillon}, {Perrier}, {Udry}, {Bouchy}, {Lovis}, {Pepe}, {Queloz}, {Santos},
  \& {Bertaux}}]{2007A&A...474..293B}
{Bonfils}, X., {Mayor}, M., {Delfosse}, X., {et~al.} 2007, \aap, 474, 293

\bibitem[{{Bonfils, X.} {et~al.}(2013){Bonfils, X.}, {Delfosse, X.}, {Udry,
  S.}, {Forveille, T.}, {Mayor, M.}, {Perrier, C.}, {Bouchy, F.}, {Gillon, M.},
  {Lovis, C.}, {Pepe, F.}, {Queloz, D.}, {Santos, N. C.}, {S\'egransan, D.}, \&
  {Bertaux, J.-L.}}]{Bonfils_2013}
{Bonfils, X.}, {Delfosse, X.}, {Udry, S.}, {et~al.} 2013, A\&A, 549, A109

\bibitem[{{Bouchy} \& {Doyon}(2018)}]{2018EPSC...12.1147B}
{Bouchy}, F. \& {Doyon}, R. 2018, in European Planetary Science Congress,
  EPSC2018--1147

\bibitem[{{Bouchy} {et~al.}(2001){Bouchy}, {Pepe}, \&
  {Queloz}}]{2001A&A...374..733B}
{Bouchy}, F., {Pepe}, F., \& {Queloz}, D. 2001, \aap, 374, 733

\bibitem[{{Butler} {et~al.}(2017){Butler}, {Vogt}, {Laughlin}, {Burt},
  {Rivera}, {Tuomi}, {Teske}, {Arriagada}, {Diaz}, {Holden}, \&
  {Keiser}}]{2017AJ....153..208B}
{Butler}, R.~P., {Vogt}, S.~S., {Laughlin}, G., {et~al.} 2017, \aj, 153, 208

\bibitem[{{Cadieux} {et~al.}(2022){Cadieux}, {Doyon}, {Plotnykov},
  {H{\'e}brard}, {Jahandar}, {Artigau}, {Valencia}, {Cook}, {Martioli},
  {Vandal}, {Donati}, {Cloutier}, {Narita}, {Fukui}, {Hirano}, {Bouchy},
  {Cowan}, {Gonzales}, {Ciardi}, {Stassun}, {Arnold}, {Benneke}, {Boisse},
  {Bonfils}, {Carmona}, {Cort{\'e}s-Zuleta}, {Delfosse}, {Forveille},
  {Fouqu{\'e}}, {Gomes da Silva}, {Jenkins}, {Kiefer}, {K{\'o}sp{\'a}l},
  {Lafreni{\`e}re}, {Martins}, {Moutou}, {do Nascimento}, {Ould-Elhkim},
  {Pelletier}, {Twicken}, {Bouma}, {Cartwright}, {Darveau-Bernier}, {Grankin},
  {Ikoma}, {Kagetani}, {Kawauchi}, {Kodama}, {Kotani}, {Latham}, {Menou},
  {Ricker}, {Seager}, {Tamura}, {Vanderspek}, \&
  {Watanabe}}]{2022AJ....164...96C}
{Cadieux}, C., {Doyon}, R., {Plotnykov}, M., {et~al.} 2022, \aj, 164, 96

\bibitem[{{Camacho} {et~al.}(2023){Camacho}, {Faria}, \&
  {Viana}}]{2023MNRAS.519.5439C}
{Camacho}, J.~D., {Faria}, J.~P., \& {Viana}, P.~T.~P. 2023, \mnras, 519, 5439

\bibitem[{{Challita} {et~al.}(2018){Challita}, {Reshetov}, {Baratchart},
  {Barrick}, {Carmona}, {Donati}, {Micheau}, {Moutou}, {Vall{\'e}e}, {Belot},
  {Gallou}, {Par{\`e}s}, {Rabou}, {Thibault}, {Kouach}, {Lacombe},
  {Saddlemyer}, \& {Striebig}}]{2018SPIE10702E..62C}
{Challita}, Z., {Reshetov}, V., {Baratchart}, S., {et~al.} 2018, in Society of
  Photo-Optical Instrumentation Engineers (SPIE) Conference Series, Vol. 10702,
  Ground-based and Airborne Instrumentation for Astronomy VII, ed. C.~J.
  {Evans}, L.~{Simard}, \& H.~{Takami}, 1070262

\bibitem[{{Collier Cameron} {et~al.}(2021){Collier Cameron}, {Ford}, {Shahaf},
  {Aigrain}, {Dumusque}, {Haywood}, {Mortier}, {Phillips}, {Buchhave},
  {Cecconi}, {Cegla}, {Cosentino}, {Cr{\'e}tignier}, {Ghedina}, {Gonz{\'a}lez},
  {Latham}, {Lodi}, {L{\'o}pez-Morales}, {Micela}, {Molinari}, {Pepe},
  {Piotto}, {Poretti}, {Queloz}, {Juan}, {S{\'e}gransan}, {Sozzetti},
  {Szentgyorgyi}, {Thompson}, {Udry}, \& {Watson}}]{2021MNRAS.505.1699C}
{Collier Cameron}, A., {Ford}, E.~B., {Shahaf}, S., {et~al.} 2021, \mnras, 505,
  1699

\bibitem[{{Cook} {et~al.}(2022){Cook}, {Hobson}, {Bouchy}, \&
  {Martioli}}]{2022ascl.soft11019C}
{Cook}, N., {Hobson}, M., {Bouchy}, F., \& {Martioli}, E. 2022, {APERO: A
  PipelinE to Reduce Observations}, Astrophysics Source Code Library, record
  ascl:2211.019

\bibitem[{{Courcol} {et~al.}(2015){Courcol}, {Bouchy}, {Pepe}, {Santerne},
  {Delfosse}, {Arnold}, {Astudillo-Defru}, {Boisse}, {Bonfils}, {Borgniet},
  {Bourrier}, {Cabrera}, {Deleuil}, {Demangeon}, {D{\'\i}az}, {Ehrenreich},
  {Forveille}, {H{\'e}brard}, {Lagrange}, {Montagnier}, {Moutou}, {Rey},
  {Santos}, {S{\'e}gransan}, {Udry}, \& {Wilson}}]{2015A&A...581A..38C}
{Courcol}, B., {Bouchy}, F., {Pepe}, F., {et~al.} 2015, \aap, 581, A38

\bibitem[{{Cretignier} {et~al.}(2020){Cretignier}, {Dumusque}, {Allart},
  {Pepe}, \& {Lovis}}]{Cretignier_2020}
{Cretignier}, M., {Dumusque}, X., {Allart}, R., {Pepe}, F., \& {Lovis}, C.
  2020, Astronomy and Astrophysics, 633, A76

\bibitem[{{Cretignier} {et~al.}(2021){Cretignier}, {Dumusque}, {Hara}, \&
  {Pepe}}]{2021A&A...653A..43C}
{Cretignier}, M., {Dumusque}, X., {Hara}, N.~C., \& {Pepe}, F. 2021, \aap, 653,
  A43

\bibitem[{{Cretignier} {et~al.}(2022){Cretignier}, {Dumusque}, \&
  {Pepe}}]{2022A&A...659A..68C}
{Cretignier}, M., {Dumusque}, X., \& {Pepe}, F. 2022, \aap, 659, A68

\bibitem[{{Cristofari} {et~al.}(2022){Cristofari}, {Donati}, {Masseron},
  {Fouqu{\'e}}, {Moutou}, {Carmona}, {Artigau}, {Martioli}, {H{\'e}brard},
  {Gaidos}, {Delfosse}, \& {SLS consortium}}]{2022MNRAS.516.3802C}
{Cristofari}, P.~I., {Donati}, J.~F., {Masseron}, T., {et~al.} 2022, \mnras,
  516, 3802

\bibitem[{da~Silva {et~al.}(2012)da~Silva, Santos, Bonfils, Delfosse,
  Forveille, Udry, Dumusque, \& Lovis}]{Gomes_da_Silva_2012}
da~Silva, J.~G., Santos, N.~C., Bonfils, X., {et~al.} 2012, Astronomy and
  Astrophysics, 541, A9

\bibitem[{{Delchambre}(2015)}]{2015MNRAS.446.3545D}
{Delchambre}, L. 2015, \mnras, 446, 3545

\bibitem[{{Donati} {et~al.}(2020){Donati}, {Kouach}, {Moutou}, {Doyon},
  {Delfosse}, {Artigau}, {Baratchart}, {Lacombe}, {Barrick}, {H{\'e}brard},
  {Bouchy}, {Saddlemyer}, {Par{\`e}s}, {Rabou}, {Micheau}, {Dolon}, {Reshetov},
  {Challita}, {Carmona}, {Striebig}, {Thibault}, {Martioli}, {Cook},
  {Fouqu{\'e}}, {Vermeulen}, {Wang}, {Arnold}, {Pepe}, {Boisse}, {Figueira},
  {Bouvier}, {Ray}, {Feugeade}, {Morin}, {Alencar}, {Hobson}, {Castilho},
  {Udry}, {Santos}, {Hernandez}, {Benedict}, {Vall{\'e}e}, {Gallou}, {Dupieux},
  {Larrieu}, {Perruchot}, {Sottile}, {Moreau}, {Usher}, {Baril}, {Wildi},
  {Chazelas}, {Malo}, {Bonfils}, {Loop}, {Kerley}, {Wevers}, {Dunn}, {Pazder},
  {Macdonald}, {Dubois}, {Carri{\'e}}, {Valentin}, {Henault}, {Yan}, \&
  {Steinmetz}}]{2020MNRAS.498.5684D}
{Donati}, J.~F., {Kouach}, D., {Moutou}, C., {et~al.} 2020, \mnras, 498, 5684

\bibitem[{{Dressing} \& {Charbonneau}(2015)}]{DressingCharbonneau2015}
{Dressing}, C.~D. \& {Charbonneau}, D. 2015, \apj, 807, 45

\bibitem[{Dumusque(2018)}]{Dumusque_2018}
Dumusque, X. 2018, Astronomy and Astrophysics, 620, A47

\bibitem[{{Dumusque} {et~al.}(2015){Dumusque}, {Pepe}, {Lovis}, \&
  {Latham}}]{2015ApJ...808..171D}
{Dumusque}, X., {Pepe}, F., {Lovis}, C., \& {Latham}, D.~W. 2015, \apj, 808,
  171

\bibitem[{Faria {et~al.}(2022)Faria, Mascare{\~{n}}o, Figueira, Silva, Damasso,
  Demangeon, Pepe, Santos, Rebolo, Cristiani, Adibekyan, Alibert, Allart,
  Barros, Cabral, D'Odorico, Marcantonio, Dumusque, Ehrenreich,
  Hern{\'{a}}ndez, Hara, Lillo-Box, Curto, Lovis, Martins, M{\'{e}}gevand,
  Mehner, Micela, Molaro, Nunes, Pall{\'{e}}, Poretti, Sousa, Sozzetti,
  Tabernero, Udry, \& Osorio}]{Faria_2022}
Faria, J.~P., Mascare{\~{n}}o, A.~S., Figueira, P., {et~al.} 2022, Astronomy
  and Astrophysics, 658, A115

\bibitem[{{Foreman-Mackey} {et~al.}(2017){Foreman-Mackey}, {Agol}, {Angus}, \&
  {Ambikasaran}}]{celerite}
{Foreman-Mackey}, D., {Agol}, E., {Angus}, R., \& {Ambikasaran}, S. 2017, ArXiv

\bibitem[{{Foreman-Mackey} {et~al.}(2013){Foreman-Mackey}, {Conley},
  {Meierjurgen Farr}, {Hogg}, {Lang}, {Marshall}, {Price-Whelan}, {Sanders}, \&
  {Zuntz}}]{2013ascl.soft03002F}
{Foreman-Mackey}, D., {Conley}, A., {Meierjurgen Farr}, W., {et~al.} 2013,
  {emcee: The MCMC Hammer}, Astrophysics Source Code Library, record
  ascl:1303.002

\bibitem[{{Fouqu{\'e}} {et~al.}(2023){Fouqu{\'e}}, {Martioli}, {Donati},
  {Lehmann}, {Zaire}, {Bellotti}, {Gaidos}, {Morin}, {Moutou}, {Petit},
  {Alencar}, {Arnold}, {Artigau}, {Cang}, {Carmona}, {Cook},
  {Cort{\'e}s-Zuleta}, {Cristofari}, {Delfosse}, {Doyon}, {H{\'e}brard},
  {Malo}, {Reyl{\'e}}, \& {Usher}}]{2023A&A...672A..52F}
{Fouqu{\'e}}, P., {Martioli}, E., {Donati}, J.~F., {et~al.} 2023, \aap, 672,
  A52

\bibitem[{{Fulton} {et~al.}(2018){Fulton}, {Petigura}, {Blunt}, \&
  {Sinukoff}}]{2018PASP..130d4504F}
{Fulton}, B.~J., {Petigura}, E.~A., {Blunt}, S., \& {Sinukoff}, E. 2018, \pasp,
  130, 044504

\bibitem[{{Gaia Collaboration}(2020)}]{2020yCat.1350....0G}
{Gaia Collaboration}. 2020, VizieR Online Data Catalog, I/350

\bibitem[{{Gaidos} {et~al.}(2016){Gaidos}, {Mann}, {Kraus}, \&
  {Ireland}}]{2016MNRAS.457.2877G}
{Gaidos}, E., {Mann}, A.~W., {Kraus}, A.~L., \& {Ireland}, M. 2016, \mnras,
  457, 2877

\bibitem[{{Gibson} {et~al.}(2020){Gibson}, {Merritt}, {Nugroho}, {Cubillos},
  {de Mooij}, {Mikal-Evans}, {Fossati}, {Lothringer}, {Nikolov}, {Sing},
  {Spake}, {Watson}, \& {Wilson}}]{2020MNRAS.493.2215G}
{Gibson}, N.~P., {Merritt}, S., {Nugroho}, S.~K., {et~al.} 2020, \mnras, 493,
  2215

\bibitem[{{H{\'e}brard} {et~al.}(2016){H{\'e}brard}, {Donati}, {Delfosse},
  {Morin}, {Moutou}, \& {Boisse}}]{2016MNRAS.461.1465H}
{H{\'e}brard}, {\'E}.~M., {Donati}, J.~F., {Delfosse}, X., {et~al.} 2016,
  \mnras, 461, 1465

\bibitem[{{Henry} {et~al.}(2006){Henry}, {Jao}, {Subasavage}, {Beaulieu},
  {Ianna}, {Costa}, \& {M{\'e}ndez}}]{2006AJ....132.2360H}
{Henry}, T.~J., {Jao}, W.-C., {Subasavage}, J.~P., {et~al.} 2006, \aj, 132,
  2360

\bibitem[{{Hobson} {et~al.}(2021){Hobson}, {Bouchy}, {Cook}, {Artigau},
  {Moutou}, {Boisse}, {Lovis}, {Carmona}, {Delfosse}, {Donati}, \& {SPIRou
  Team}}]{2021A&A...648A..48H}
{Hobson}, M.~J., {Bouchy}, F., {Cook}, N.~J., {et~al.} 2021, \aap, 648, A48

\bibitem[{{Hsu} {et~al.}(2020){Hsu}, {Ford}, \&
  {Terrien}}]{2020MNRAS.498.2249H}
{Hsu}, D.~C., {Ford}, E.~B., \& {Terrien}, R. 2020, \mnras, 498, 2249

\bibitem[{{Kanodia} \& {Wright}(2018)}]{2018RNAAS...2....4K}
{Kanodia}, S. \& {Wright}, J. 2018, Research Notes of the American Astronomical
  Society, 2, 4

\bibitem[{{Kingma} \& {Welling}(2013)}]{2013arXiv1312.6114K}
{Kingma}, D.~P. \& {Welling}, M. 2013, arXiv e-prints, arXiv:1312.6114

\bibitem[{{Kirkpatrick} {et~al.}(1991){Kirkpatrick}, {Henry}, \&
  {McCarthy}}]{1991ApJS...77..417K}
{Kirkpatrick}, J.~D., {Henry}, T.~J., \& {McCarthy}, Donald~W., J. 1991, \apjs,
  77, 417

\bibitem[{{Kotani} {et~al.}(2014){Kotani}, {Tamura}, {Suto}, {Nishikawa},
  {Sato}, {Aoki}, {Usuda}, {Kurokawa}, {Kashiwagi}, {Nishiyama}, {Ikeda},
  {Hall}, {Hodapp}, {Hashimoto}, {Morino}, {Okuyama}, {Tanaka}, {Suzuki},
  {Inoue}, {Kwon}, {Suenaga}, {Oh}, {Baba}, {Narita}, {Kokubo}, {Hayano},
  {Izumiura}, {Kambe}, {Kudo}, {Kusakabe}, {Ikoma}, {Hori}, {Omiya}, {Genda},
  {Fukui}, {Fujii}, {Guyon}, {Harakawa}, {Hayashi}, {Hidai}, {Hirano},
  {Kuzuhara}, {Machida}, {Matsuo}, {Nagata}, {Onuki}, {Ogihara}, {Takami},
  {Takato}, {Takahashi}, {Tachinami}, {Terada}, {Kawahara}, \&
  {Yamamuro}}]{2014SPIE.9147E..14K}
{Kotani}, T., {Tamura}, M., {Suto}, H., {et~al.} 2014, in Society of
  Photo-Optical Instrumentation Engineers (SPIE) Conference Series, Vol. 9147,
  Ground-based and Airborne Instrumentation for Astronomy V, ed. S.~K.
  {Ramsay}, I.~S. {McLean}, \& H.~{Takami}, 914714

\bibitem[{{Mahadevan} {et~al.}(2012){Mahadevan}, {Ramsey}, {Bender}, {Terrien},
  {Wright}, {Halverson}, {Hearty}, {Nelson}, {Burton}, {Redman}, {Osterman},
  {Diddams}, {Kasting}, {Endl}, \& {Deshpande}}]{2012SPIE.8446E..1SM}
{Mahadevan}, S., {Ramsey}, L., {Bender}, C., {et~al.} 2012, in Society of
  Photo-Optical Instrumentation Engineers (SPIE) Conference Series, Vol. 8446,
  Ground-based and Airborne Instrumentation for Astronomy IV, ed. I.~S.
  {McLean}, S.~K. {Ramsay}, \& H.~{Takami}, 84461S

\bibitem[{{Martioli} {et~al.}(2022){Martioli}, {H{\'e}brard}, {Fouqu{\'e}},
  {Artigau}, {Donati}, {Cadieux}, {Bellotti}, {Lecavelier des Etangs}, {Doyon},
  {do Nascimento}, {Arnold}, {Carmona}, {Cook}, {Cortes-Zuleta}, {de Almeida},
  {Delfosse}, {Folsom}, {K{\"o}nig}, {Moutou}, {Ould-Elhkim}, {Petit},
  {Stassun}, {Vidotto}, {Vandal}, {Benneke}, {Boisse}, {Bonfils}, {Boyd},
  {Brasseur}, {Charbonneau}, {Cloutier}, {Collins}, {Cristofari}, {Crossfield},
  {D{\'\i}az}, {Fausnaugh}, {Figueira}, {Forveille}, {Furlan}, {Girardin},
  {Gnilka}, {Gomes da Silva}, {Gu}, {Guerra}, {Howell}, {Hussain}, {Jenkins},
  {Kiefer}, {Latham}, {Matson}, {Matthews}, {Morin}, {Naves}, {Ricker},
  {Seager}, {Takami}, {Twicken}, {Vanderburg}, {Vanderspek}, \&
  {Winn}}]{2022A&A...660A..86M}
{Martioli}, E., {H{\'e}brard}, G., {Fouqu{\'e}}, P., {et~al.} 2022, \aap, 660,
  A86

\bibitem[{Mascare{\~{n}}o {et~al.}(2023)Mascare{\~{n}}o,
  Gonz{\'{a}}lez-{\'{A}}lvarez, Osorio, Lillo-Box, Faria, Passegger,
  Hern{\'{a}}ndez, Figueira, Sozzetti, Rebolo, Pepe, Santos, Cristiani, Lovis,
  Silva, Ribas, Amado, Caballero, Quirrenbach, Reiners, Zechmeister, Adibekyan,
  Alibert, B{\'{e}}jar, Benatti, D'Odorico, Damasso, Delisle, Marcantonio,
  Dreizler, Ehrenreich, Hatzes, Hara, Henning, Kaminski,
  L{\'{o}}pez-Gonz{\'{a}}lez, Martins, Micela, Montes, Pall{\'{e}}, Pedraz,
  Rodr{\'{\i}}guez, Rodr{\'{\i}}guez-L{\'{o}}pez, Tal-Or, Sousa, \&
  Udry}]{Su_rez_Mascare_o_2023}
Mascare{\~{n}}o, A.~S., Gonz{\'{a}}lez-{\'{A}}lvarez, E., Osorio, M. R.~Z.,
  {et~al.} 2023, Astronomy and Astrophysics, 670, A5

\bibitem[{{Micheau} {et~al.}(2018){Micheau}, {Kouach}, {Donati}, {Gallou},
  {Belot}, {Striebig}, {Baratchart}, {Challita}, {Par{\`e}s}, {Dubois},
  {Lacombe}, {Barrick}, {Bouchy}, \& {Pepe}}]{2018SPIE10702E..5RM}
{Micheau}, Y., {Kouach}, D., {Donati}, J.-F., {et~al.} 2018, in Society of
  Photo-Optical Instrumentation Engineers (SPIE) Conference Series, Vol. 10702,
  Ground-based and Airborne Instrumentation for Astronomy VII, ed. C.~J.
  {Evans}, L.~{Simard}, \& H.~{Takami}, 107025R

\bibitem[{{Newton} {et~al.}(2016){Newton}, {Irwin}, {Charbonneau},
  {Berta-Thompson}, \& {Dittmann}}]{2016ApJ...821L..19N}
{Newton}, E.~R., {Irwin}, J., {Charbonneau}, D., {Berta-Thompson}, Z.~K., \&
  {Dittmann}, J.~A. 2016, \apjl, 821, L19

\bibitem[{{Pakhomov} {et~al.}(2017){Pakhomov}, {Piskunov}, \&
  {Ryabchikova}}]{2017ASPC..510..518P}
{Pakhomov}, Y., {Piskunov}, N., \& {Ryabchikova}, T. 2017, in Astronomical
  Society of the Pacific Conference Series, Vol. 510, Stars: From Collapse to
  Collapse, ed. Y.~Y. {Balega}, D.~O. {Kudryavtsev}, I.~I. {Romanyuk}, \& I.~A.
  {Yakunin}, 518

\bibitem[{{Par{\`e}s} {et~al.}(2012){Par{\`e}s}, {Donati}, {Dupieux}, {Gharsa},
  {Micheau}, {Bouye}, {Dubois}, {Gallou}, {Kouach}, {Barrick}, \&
  {Wang}}]{2012SPIE.8446E..2EP}
{Par{\`e}s}, L., {Donati}, J.~F., {Dupieux}, M., {et~al.} 2012, in Society of
  Photo-Optical Instrumentation Engineers (SPIE) Conference Series, Vol. 8446,
  Ground-based and Airborne Instrumentation for Astronomy IV, ed. I.~S.
  {McLean}, S.~K. {Ramsay}, \& H.~{Takami}, 84462E

\bibitem[{{Perger} {et~al.}(2023){Perger}, {Anglada-Escud{\'e}}, {Baroch},
  {Lafarga}, {Ribas}, {Morales}, {Herrero}, {Amado}, {Barnes}, {Caballero},
  {Jeffers}, {Quirrenbach}, \& {Reiners}}]{2023A&A...672A.118P}
{Perger}, M., {Anglada-Escud{\'e}}, G., {Baroch}, D., {et~al.} 2023, \aap, 672,
  A118

\bibitem[{Quirrenbach {et~al.}(2018)Quirrenbach, Amado, Ribas, Caballero,
  Seifert, Aceituno, Azzaro, Barrado, Becerril, B{\`e}jar, Benítez,
  Brinkmöller, Colom{\'e}, Cort{\'e}s-Contreras, Czesla, Frölich,
  Galadí-Enríquez, González~Hernández, González~Peinado, Guenther,
  de~Guindos, Hagen, Henning, Hernández~Castaño, Herrero, Hintz, Jeffers,
  Kaminski, Klahr, Marfil, Martín, Martín-Ruiz, Mathar, Montes, Morales,
  Nagel, Pall{\'e}, P{\'e}rez-Medialdea, Perger, Rebolo, Reffert, Rosich,
  Sabotta, Schäfer, Schiller, Schweitzer, Solano, Stahl, Tala~Pinto, Trifonov,
  Yan, Zechmeister, Abellán, Abril, Alonso-Floriano, Ammler-von Eiff,
  Anglada-Escud{\'e}, Anwand-Heerwart, Berdiñas, Bergondy, del Burgo,
  Cárdenas, Casal, Claret, Ferro, Gálvez-Ortiz, Gesa, Gómez~Galera,
  Guijarro, Hedrosa, Hermann, Hermelo, Hernández~Arabí, Hidalgo, Huber,
  Huber, Kehr, Klein, Klüter, Klutsch, Labarga, Labiche, Lamert, Lemke,
  Lenzen, Lizon, Lodieu, López-Morales, López~Salas, López-Santiago,
  Martínez-Rodríguez, Maroto~Fernández, Marvin, Mirabet, Moreno-Raya, Moya,
  Naranjo, Pascual, P{\'e}rez-Calpena, Perryman, Rohloff, Sánchez~Carrasco,
  Schmidt, Strachan, Tal-Or, Tulloch, Veredas, Vilardell, Wagner, Zhao,
  Reiners, Baroch, Bauer, Cardona~Guill{\'e}n, Cifuentes, Dreizler,
  Fuhrmeister, Hatzes, Hauschildt, Helmling, Herbort, Johnson, de~Juan,
  Kürster, Lafarga, Sairam, Lampón, Lara, Launhardt, López~del Fresno,
  López-Puertas, Luque, Mandel, Nortmann, Nowak, Passegger, Pavlov, Pedraz,
  Rodríguez, Rodríguez~López, Sadegi, Salz, Sánchez-López, Sanz-Forcada,
  Sarkis, Schmitt, Schöfer, Shulyak, Zapatero~Osorio, Arroyo-Torres, Blümcke,
  Cano, Carro, Díez-Alonso, Doellinger, Dorda, Feiz, Fernández, Gaisn{\'e},
  Gallardo, García-Piquer, García-Vargas, Garrido, González-Álvarez,
  González-Cuesta, Grohnert, Grözinger, Guàrdia, Hernández~Hernando,
  Holgado, Huke, Kim, Laun, Lázaro, Llamas, López~González,
  Magán~Madinabeitia, Mall, Mancini, Marín~Molina, Mundt, Panduro, Pluto,
  Ramón, Redondo, Reinhart, Rhode, Rix, Rodler, Sánchez-Blanco, Sarmiento,
  Storz, Stürmer, Suárez, Tabernero, Ulbrich, Vico~Linares, Vidal-Dasilva,
  Winkler, Wolthoff, \& Xu}]{quirrenbach_carmenes_2018}
Quirrenbach, A., Amado, P.~J., Ribas, I., {et~al.} 2018, in Ground-based and
  {Airborne} {Instrumentation} for {Astronomy} {VII}, ed. H.~Takami, C.~J.
  Evans, \& L.~Simard (Austin, United States: SPIE), 32

\bibitem[{{Reiners} \& {Zechmeister}(2020)}]{2020ApJS..247...11R}
{Reiners}, A. \& {Zechmeister}, M. 2020, \apjs, 247, 11

\bibitem[{{Sabotta} {et~al.}(2021){Sabotta}, {Schlecker}, {Chaturvedi},
  {Guenther}, {Mu{\~n}oz Rodr{\'\i}guez}, {Mu{\~n}oz S{\'a}nchez}, {Caballero},
  {Shan}, {Reffert}, {Ribas}, {Reiners}, {Hatzes}, {Amado}, {Klahr}, {Morales},
  {Quirrenbach}, {Henning}, {Dreizler}, {Pall{\'e}}, {Perger}, {Azzaro},
  {Jeffers}, {Kaminski}, {K{\"u}rster}, {Lafarga}, {Montes}, {Passegger}, \&
  {Zechmeister}}]{2021A&A...653A.114S}
{Sabotta}, S., {Schlecker}, M., {Chaturvedi}, P., {et~al.} 2021, \aap, 653,
  A114

\bibitem[{{Schwarz}(1978)}]{1978AnSta...6..461S}
{Schwarz}, G. 1978, Annals of Statistics, 6, 461

\bibitem[{{Stock} {et~al.}(2020){Stock}, {Nagel}, {Kemmer}, {Passegger},
  {Reffert}, {Quirrenbach}, {Caballero}, {Czesla}, {B{\'e}jar}, {Cardona},
  {D{\'\i}ez-Alonso}, {Herrero}, {Lalitha}, {Schlecker}, {Tal-Or},
  {Rodr{\'\i}guez}, {Rodr{\'\i}guez-L{\'o}pez}, {Ribas}, {Reiners}, {Amado},
  {Bauer}, {Bluhm}, {Cort{\'e}s-Contreras}, {Gonz{\'a}lez-Cuesta}, {Dreizler},
  {Hatzes}, {Henning}, {Jeffers}, {Kaminski}, {K{\"u}rster}, {Lafarga},
  {L{\'o}pez-Gonz{\'a}lez}, {Montes}, {Morales}, {Pedraz}, {Sch{\"o}fer},
  {Schweitzer}, {Trifonov}, {Zapatero Osorio}, \&
  {Zechmeister}}]{2020A&A...643A.112S}
{Stock}, S., {Nagel}, E., {Kemmer}, J., {et~al.} 2020, \aap, 643, A112

\bibitem[{{Tal-Or} {et~al.}(2019){Tal-Or}, {Trifonov}, {Zucker}, {Mazeh}, \&
  {Zechmeister}}]{2019MNRAS.484L...8T}
{Tal-Or}, L., {Trifonov}, T., {Zucker}, S., {Mazeh}, T., \& {Zechmeister}, M.
  2019, \mnras, 484, L8

\bibitem[{{Tamuz} {et~al.}(2005){Tamuz}, {Mazeh}, \&
  {Zucker}}]{2005MNRAS.356.1466T}
{Tamuz}, O., {Mazeh}, T., \& {Zucker}, S. 2005, \mnras, 356, 1466

\bibitem[{{The Astropy Collaboration}(2018)}]{2018zndo...4080996T}
{The Astropy Collaboration}. 2018, {astropy v3.1: a core python package for
  astronomy}, Zenodo

\bibitem[{{Trifonov} {et~al.}(2020){Trifonov}, {Tal-Or}, {Zechmeister},
  {Kaminski}, {Zucker}, \& {Mazeh}}]{2020A&A...636A..74T}
{Trifonov}, T., {Tal-Or}, L., {Zechmeister}, M., {et~al.} 2020, \aap, 636, A74

\bibitem[{Vieira(2012)}]{permutation_test}
Vieira, V. 2012, Computational Ecology and Software, 2, 103

\bibitem[{Zechmeister \& Kürster(2009)}]{Zechmeister_2009}
Zechmeister, M. \& Kürster, M. 2009, Astronomy and Astrophysics, 496, 577

\bibitem[{{Zechmeister} {et~al.}(2018){Zechmeister}, {Reiners}, {Amado},
  {Azzaro}, {Bauer}, {B{\'e}jar}, {Caballero}, {Guenther}, {Hagen}, {Jeffers},
  {Kaminski}, {K{\"u}rster}, {Launhardt}, {Montes}, {Morales}, {Quirrenbach},
  {Reffert}, {Ribas}, {Seifert}, {Tal-Or}, \& {Wolthoff}}]{2018A&A...609A..12Z}
{Zechmeister}, M., {Reiners}, A., {Amado}, P.~J., {et~al.} 2018, \aap, 609, A12

\end{thebibliography}

\newpage

\begin{appendix}

\section{Supplementary information regarding the outliers rejection}

\label{determiningkmad}

The process of identifying outliers for rejection involved using the $D$ metric which quantifies the degree of anomaly of an epoch. They were subsequently ranked in decreasing order of $D$ and were removed one by one, and we stop when $D \leq 10$. The \texttt{Wapiti} method was applied to the resulting time series and the one with the maximum statistical significance of the highest peak in the resulting periodogram was chosen, which in our case resulted in the rejection of 7 outliers. This significance was determined using the log false alarm probability (FAP), but we also verified that the same conclusion was reached using the log Bayes factor.

\par 

\begin{figure}[!h]
    \centering
    \includegraphics[width=\linewidth]{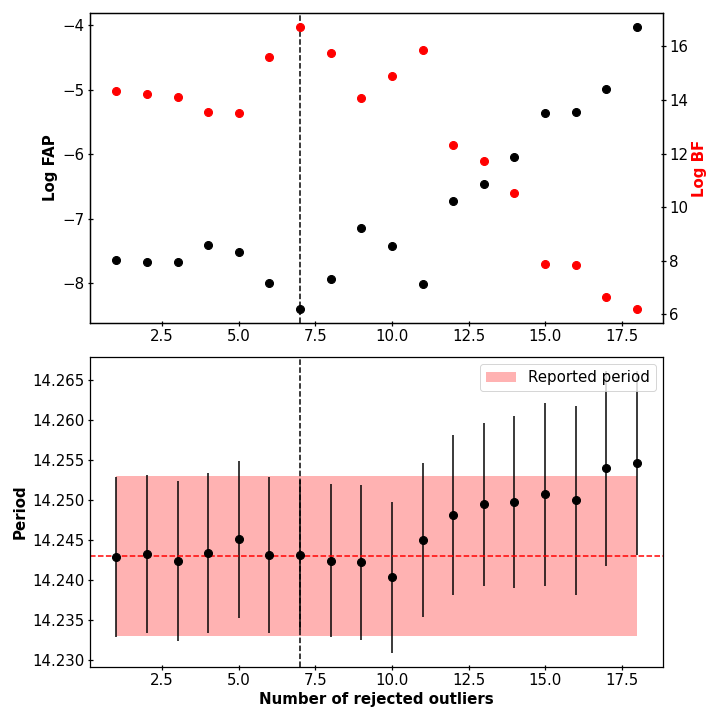}
    \caption{The top panel displays the evolution of the log FAP of the signal as the number of outliers removed increases. The red dots overlaid on the plot represent the evolution of the log BF, which follows a similar pattern and reaches its maximum value at the same number of rejected outliers. The bottom panel of the figure shows the detected period in the resulting time-series, which remains consistent within error-bars with the reported planetary period.}
    \label{fig:kmadchoice}
\end{figure}

Such a criterion to remove outliers is justified for our data set from the observation that the \texttt{Wapiti} method yields a prominent peak at the same period independently of the number of rejected outliers as shown in Figure \ref{fig:kmadchoice}. From the definition of $D$ it comes that those rejected epochs are outliers in at least one principal component, yet they do not visually appear in the RV time-series and one may be interested in what could be their origin.

\par

\begin{figure*}[!h]
    \centering
    \includegraphics[width=\linewidth]{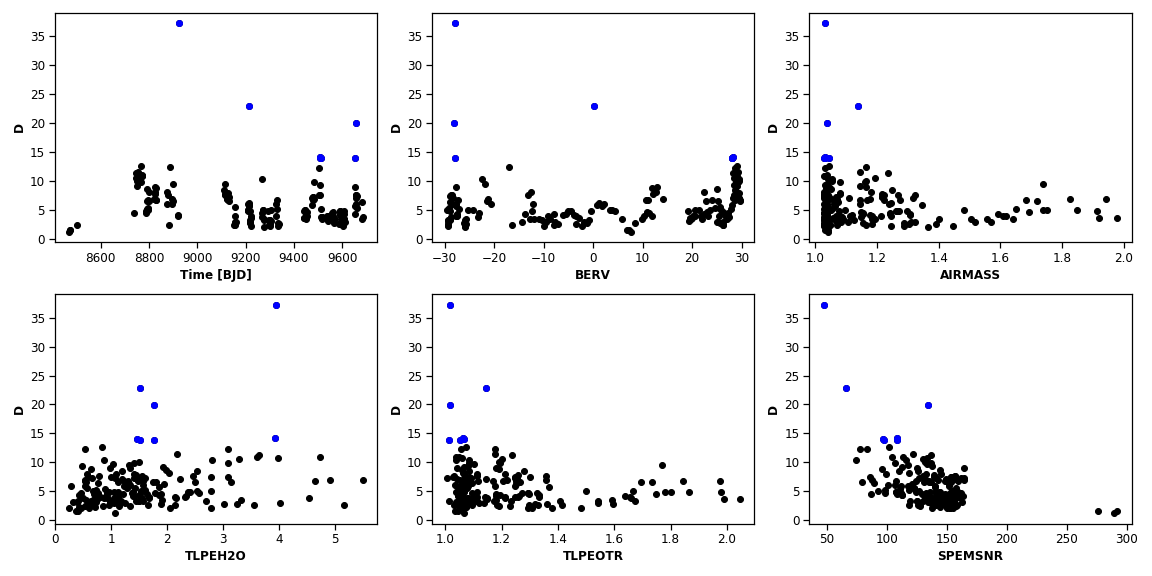}
    \caption{All panels display the degree of anomaly $D$ of the observations with respect to various parameters. The data points that appear in blue represent the rejected observations that were excluded to maximize the significance of the planetary signal in our data.}
    \label{fig:outliers_param}
\end{figure*}

This is the reason why we tried to visualize their physical origin in Figure \ref{fig:outliers_param} using the most-straightforward parameters available from the \texttt{APERO} reduction, namely the BERV, the airmass (AIRMASS), the SPIRou Exposure Meter SNR Estimate (SPEMSNR), and the telluric water exponent (TLPEH2O) and other telluric molecules (TLPEOTR) calculated with \texttt{TAPAS} during the precleaning of the telluric correction. Regrettably, our investigation did not provide conclusive evidence to determine the source of the rejected outliers. These outliers exhibit a wide range of BERV values and are present at low airmasses ($\leq 1.2$). Furthermore, some of the rejected outliers have a SPEMSNR that is not necessarily lower than the average, even though we do notice that the first two outliers are the data points with the lowest SPEMSNR. Ultimately, it is likely that multiple factors contribute to the anomalous behavior of these observations, and a single parameter is insufficient to explain this anomaly. We advise that when applying this outlier removal method to other targets, it should be done with caution and the results should be carefully verified, as shown in Figure \ref{fig:kmadchoice}.

\clearpage

\section{Impact of the components on the signal}
\label{impact_component}
\begin{figure}[!h]
    \centering
    \includegraphics[width=\linewidth]{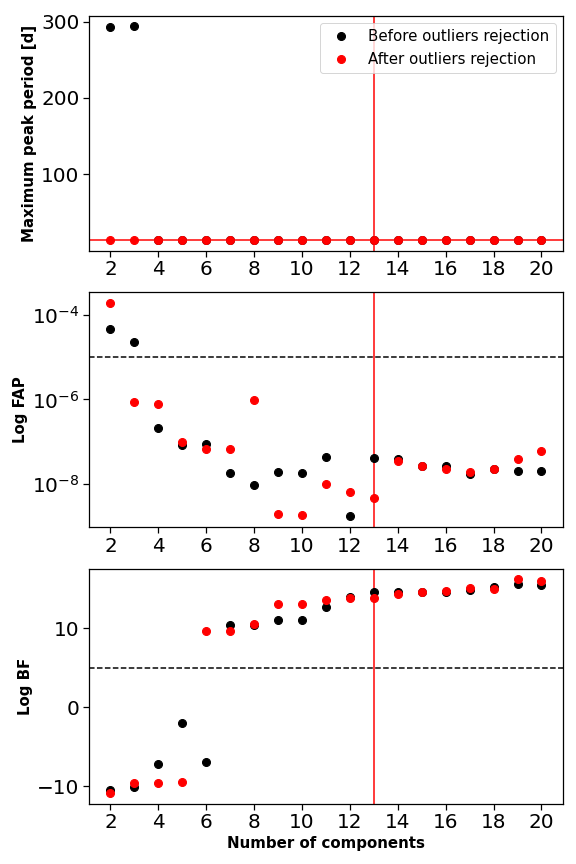}
    \caption{The top panel displays the peak associated with the maximum period in the periodogram when the wPCA reconstruction is eliminated from the data using varying numbers of components. In the middle panel, the logarithm of the FAP linked to this signal is presented, while the bottom panel depicts the $\log BF$ evaluating the significance of a 14.2\,d Keplerian signal compared to an offset. The black and red data points correspond, respectively, to the outcomes obtained from the pre- and post-outlier rejection RV time-series.}
    \label{fig:component_impact}
\end{figure}

After determining that 13 components were required to accurately reconstruct our data using wPCA, one may wonder how the number of components affects the signals reported in our data. To address this question, we conducted a straightforward test by applying the method to time-series with an increasing number of components ranging from $2$ to $20$. For each time-series, we computed its periodogram and the FAP of its maximum peak period. Additionally, we performed a computation of the $\log BF$ to compare a Keplerian model with a 14.2-day orbital period and an offset with a model solely consisting of an offset. The outcomes are exhibited in Figure \ref{fig:component_impact}. As our rejection of outliers is expected to favor the use of 13 components, we conducted this examination on both RV time-series, both before and after outliers rejection.

\par 

Contrary to expectations, the use of 13 components did not yield the highest level of statistical significance for the 14.2-day signal, but instead required 12 components without outlier rejections and 10 components with such rejections. It is noteworthy that the planetary signal can be directly extracted using only 2 components in the time-series data that underwent $\texttt{Wapiti}$ correction and outlier rejection, compared to 3 components required without such rejection. Notably, in the case of no outliers rejection, the number of components had minimal impact on the results after reaching  5 components, with the signal remaining above a $\log BF$ of 5 and a FAP value below $10^{-5}$.

\par 

To sum up, the results indicate that the number of components employed does not significantly impact the outcomes beyond a certain threshold. However, given that the objective is to use the $\texttt{Wapiti}$ method on unidentified planetary systems where the orbital period is not a priori known, an unbiased and robust approach is necessary to determine the appropriate number of components. The identification of 13 as the optimal number of components through our independent pre-processing methods provides robust evidence of the effectiveness of our methodology.

\clearpage

\section{MCMC results from the RV analysis}

Orbital parameters computed using an MCMC over the \texttt{Wapiti} corrected RV data from SPIRou with 100,000 samples, 10,000 burn-in samples, and 100 walkers. 

\begin{figure*}[!h]
    \centering
    \includegraphics[width=\linewidth]{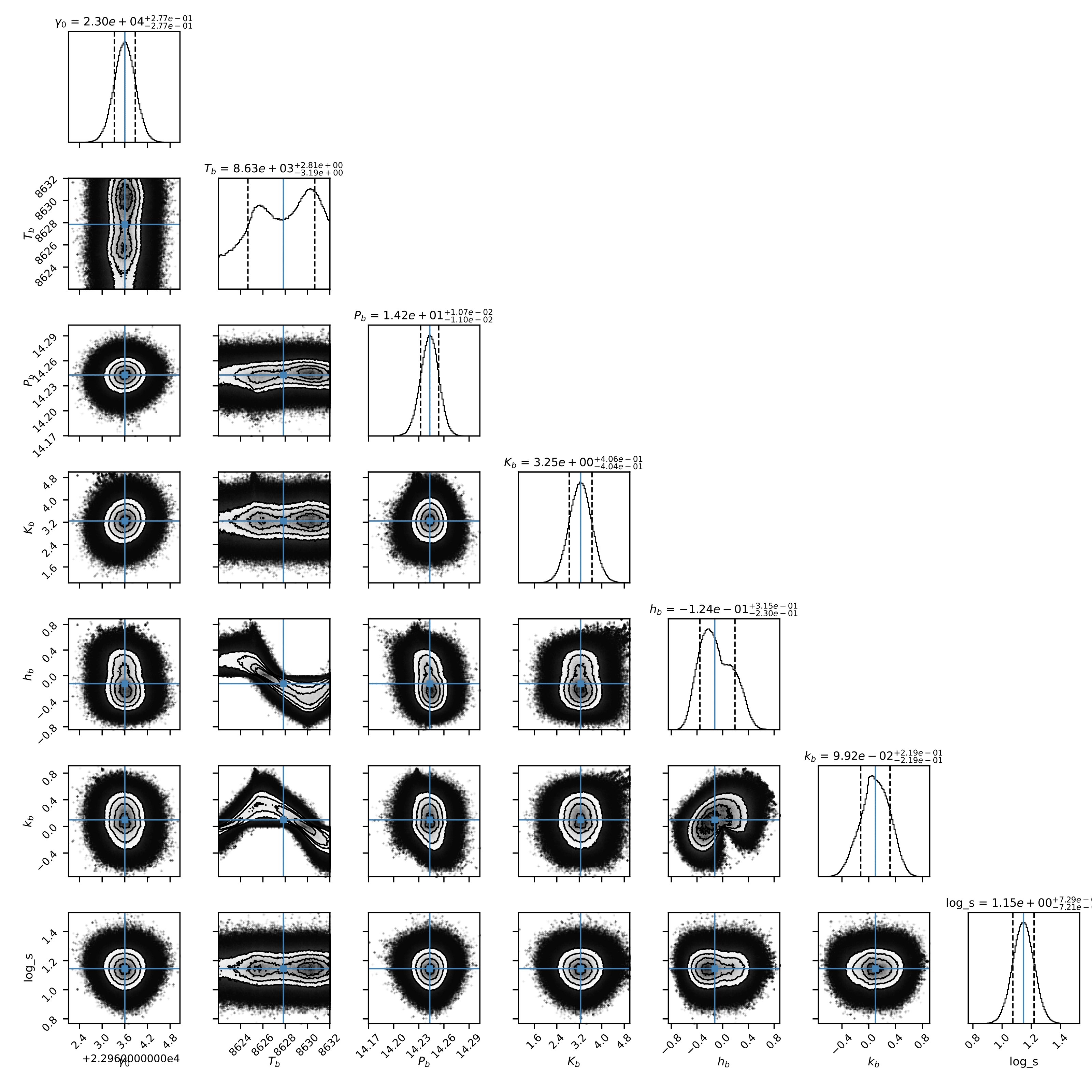}
    \caption{Corner plot displaying the results of MCMC analysis on the \texttt{Wapiti} corrected RV time-series is presented. This analysis involves fitting a model that consists of an offset $\gamma_0$, an RV jitter $s$ (in log), and a Keplerian with parameters including the time of periastron $T_b$, the orbital period $P_b$, the semi-amplitude $K_b$, and the eccentricity and argument of periastron represented by $\sqrt{e}\cos \omega$ ($h_b$) and $\sqrt{e}\sin \omega$ ($k_b$), respectively.}
    \label{fig:corner_plot}
\end{figure*}

\clearpage

\section{MCMC results from the W1 analysis}
\label{W1_analysis}

\begin{figure*}[!h]
    \centering
    \includegraphics[width=\linewidth]{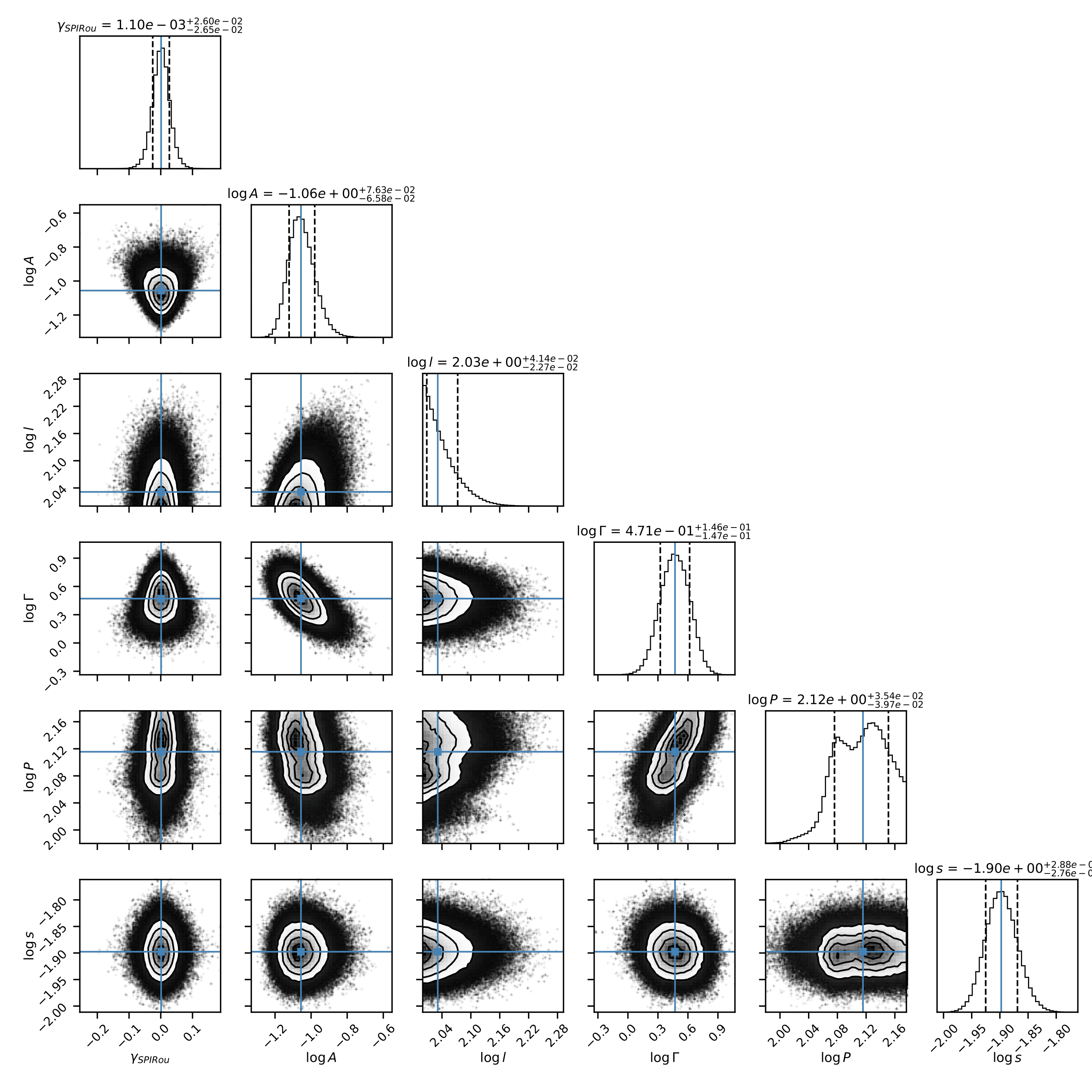}
    \caption{Corner plot displaying the results of MCMC analysis on the $W_1$ time-series. This analysis involves fitting a quasi-periodic GP over the data and an offset.}
    \label{fig:corner_plot_activity}
\end{figure*}

We fitted a quasi-periodic GP in order to estimate the value of the periodicity in $W_1$ fitting a quasi-periodic kernel of the form
\begin{equation}
    k\left(\tau\right) = A \exp\left(- \frac{\tau^2}{2l^2} - \Gamma \sin^2\left(\frac{\pi}{P}\tau \right)\right) 
\end{equation}
as well as a jitter term $s$ and a constant $\gamma$. We computed an MCMC using 10,000 steps, 1,000 burn-in samples, and 100 walkers. The result is displayed in Table \ref{GP_Table} and the corner plot is shown in Figure \ref{fig:corner_plot_activity}.

\begin{table*}[]
\centering
  \caption[]{Prior and posterior distributions of the quasi-periodic GP model of the SPIRou $W_1$ time-series.}
  \label{GP_Table}
 \begin{tabular}{llllll}
  \hline
  \hline
  \multicolumn{1}{c}{\rm Parameter} & \multicolumn{2}{l}{\rm Prior distribution} & \multicolumn{2}{c}{\rm Posterior distribution} \\
   & & & ln units & linear units \\
   \hline 
   \noalign{\smallskip}
    GP $W_1$\\
    \hline
    Amplitude                     & $\log{A}$         & $\mathcal{U}\left(-3,3\right)$                                             & $-1.07^{+0.08}_{-0.07}$          & $0.09^{+0.02}_{-0.01}$              \\[1mm]
    Decay time [days]            & $\log{l}$         & $\mathcal{U}\left(2,3\right)$                                             & $2.03^{+0.04}_{-0.02}$          & $107^{+11}_{-5}$            \\[1mm]
    Smoothing parameter           & $\log{\Gamma}$    & $\mathcal{U}\left(-5,5\right)$                                            & $0.55^{+0.25}_{-0.19}$         & $3.5^{+2.7}_{-1.3}$          \\[1mm]
    Cycle length (period) ~[days] & $\log{P}$         & $\mathcal{U}\left(\ln{\left(80\right)},\ln{\left(200\right)}\right)$   & $2.14^{+0.10}_{-0.05}$ & $137^{+37}_{-16}$ \\[1mm]
    Uncorr. Noise                & $\log{\sigma}$    & $\mathcal{U}\left(-5,0\right)$                                          & $-1.90^{+0.03}_{-0.03}$         & $0.0127^{+0.0009}_{-0.0008}$        \\[1mm] 
    \hline
 \end{tabular}
\end{table*}

\end{appendix}
\end{document}